\documentclass[aps,prd,twocolumn,10pt,groupedaddress]{revtex4-1}
\usepackage{amssymb}
\usepackage{graphicx}
\usepackage{amsmath}
\usepackage{hyperref}
\usepackage{subfigure}
\usepackage{float}
\usepackage{color}
\usepackage{ulem}
\usepackage[utf8]{inputenc}

\begin{document}

\title{Occurrence of semiclassical vacuum decay}
\author{Shao-Jiang Wang}
\email{schwang@cosmos.phy.tufts.edu}
\affiliation{Tufts Institute of Cosmology, Department of Physics and Astronomy, Tufts University, 574 Boston Avenue, Medford, Massachusetts 02155, USA}
%\date{\today}

\begin{abstract}
Recently, a novel phenomenon is observed for vacuum decay to proceed via classically allowed dynamical evolution from initial configurations of false vacuum fluctuations. With the help of some occasionally developed large fluctuations in time derivative of a homogeneous scalar field that is initially sitting at the false vacuum, the flyover vacuum decay is proposed if the initial Gaussian profile of field velocity contains a spatial region where the field velocity is large enough to flyover the potential barrier. In this paper, we point out that, even if the initial profile of field velocity is nowhere to be large enough to classically overcome the barrier, the semiclassical vacuum decay could still be possible if the initial profile of field velocity extends over large enough region. 
\end{abstract}
\maketitle

\section{Introduction}\label{sec:introduction}

Quantum tunnelling in nonrelativistic quantum mechanics of a single particle is a distinguishing feature from the classical mechanics where surmounting a potential barrier requires large enough energy instead of quantum mechanically penetrating the potential barrier with lower energy. This feature persists when generalized into the case of relativistic field theory with multiple potential minimums, one of which is absolute minimum in both classical and (perturbative) quantum sense, while the rest of which is still classically stable but metastable by barrier penetration in quantum field theory. Although the field version of barrier penetration consists of an infinite number of degrees of freedom, the decay rate of metastable minimum could be expressed in terms of the Euclidean instanton solution \cite{Coleman:1977py,Callan:1977pt} as an analog to the WKB approximation in nonrelativistic quantum mechanics.

Although the instanton approach borrows the wisdom of classically forbidden quantum penetration under the barrier, the physical picture behind the vacuum decay in field theory resembles the bubble nucleation in statistical physics \cite{Langer:1967ax,Langer:1969bc} with quantum fluctuations replaced by thermodynamic fluctuations. Indeed, there is a cold atom analog of vacuum decay in field theory from experimental \cite{Fialko_2015,Fialko:2016ggg}, theoretical \cite{Braden:2017add}, and simulational \cite{Billam:2018pvp,Braden:2019vsw} perspectives. This inspires the request for vacuum decay in field theory via classically allowed channel, which is recently realized in lattice simulation \cite{Braden:2018tky} and formulated in real-time formalism \cite{Hertzberg:2019wgx}.

The semiclassical picture of vacuum decay introduced above could be traced back to the stochastic approach to vacuum decay in \cite{Ellis:1990bv,Linde:1991sk}, where the initial conditions for the subsequent classical evolution could be drawn from realizations of random Gaussian fields that inherit the same statistics as the quantum fluctuations \cite{Braden:2018tky}. The simplest realizations for such initial conditions of semiclassical vacuum decay is the flyover vacuum decay \cite{Haiyun,Blanco-Pillado:2019xny}, where field velocity occasionally develops a Gaussian profile large enough to classically overcome the barrier directly from a homogeneous field at false vacuum. This flyover vacuum decay can be also realized with the presence of gravity for both downward and upward transitions \cite{Blanco-Pillado:2019xny}, which has far-reaching ramification \cite{Brown:2011ry} on the swampland criteria \cite{Obied:2018sgi,Ooguri:2018wrx} (see also \cite{Danielsson:2018ztv,Garg:2018reu}): the eternal inflating regions are unavoidable even in the swampy landscapes \cite{Blanco-Pillado:2019tdf}.

To have a successful flyover vacuum decay, it is required that there should be at least a spatial region of a size larger than the minimal expansion radius, where the field velocity should also be larger than a critical velocity to classically flyover the barrier. We point out in this paper that, even if the initial profile of field velocity is below the critical velocity everywhere in the region of survey, the semiclassical vacuum decay could still occur provided that the initial profile of field velocity covers spatial region large enough. This is a surprising result since, each local degree of freedom does not have enough energy to classically overcome the barrier, the whole region could still be converted into the true vacuum along classical evolutions.

The outline of this paper is as follows: in Sec.\ref{sec:flyover}, we review the flyover vacuum decay in order to fix the convention and setup we used in the following sections; in Sec. \ref{sec:semiclas}, we scan over all possible configurations of initial profile of field velocity according to their width and height. The last section is devoted to conclusion and discussions.

\section{Flyover vacuum decay}\label{sec:flyover}

\begin{figure*}
\centering
\includegraphics[width=0.45\textwidth]{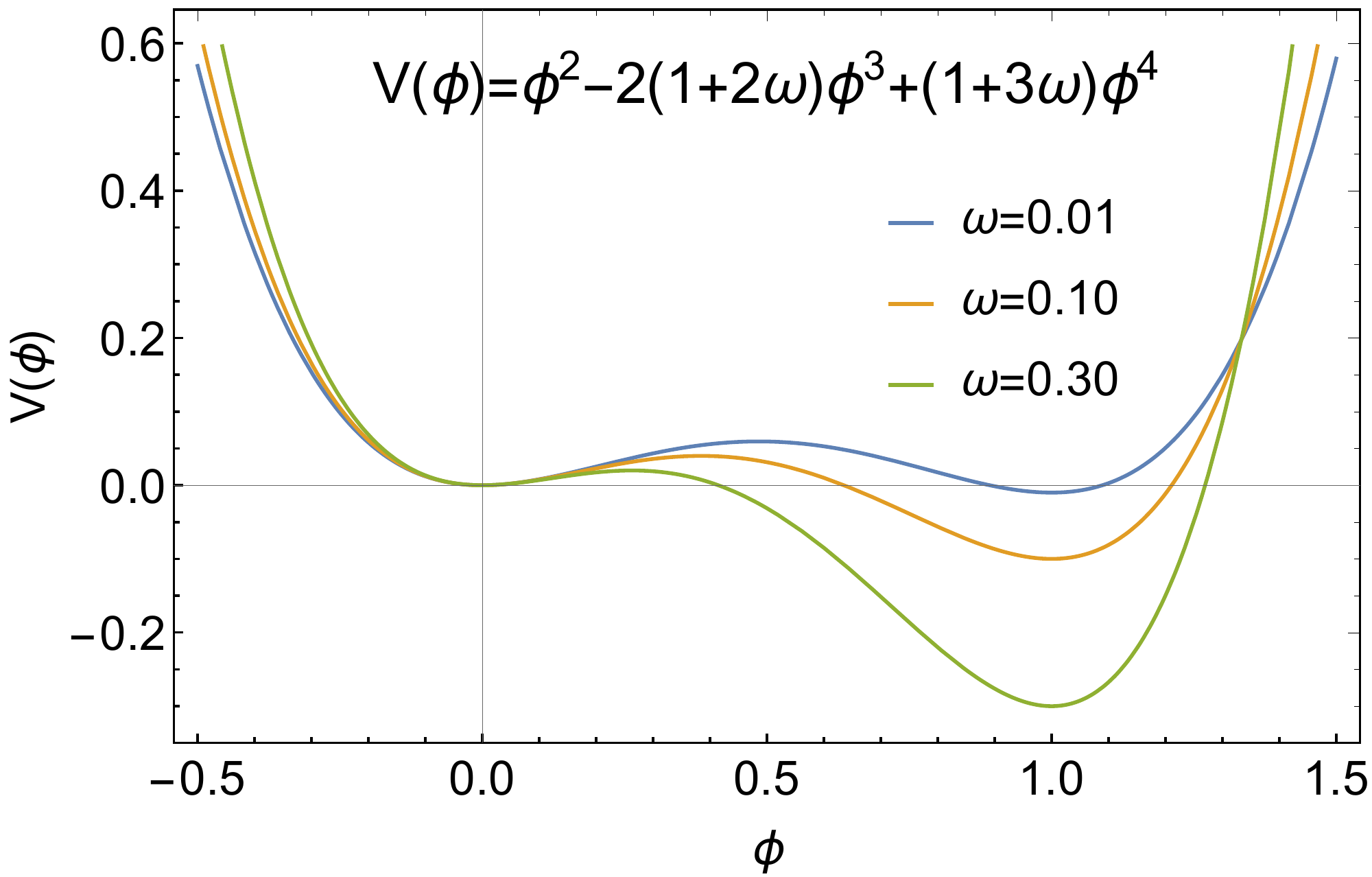}
\includegraphics[width=0.45\textwidth]{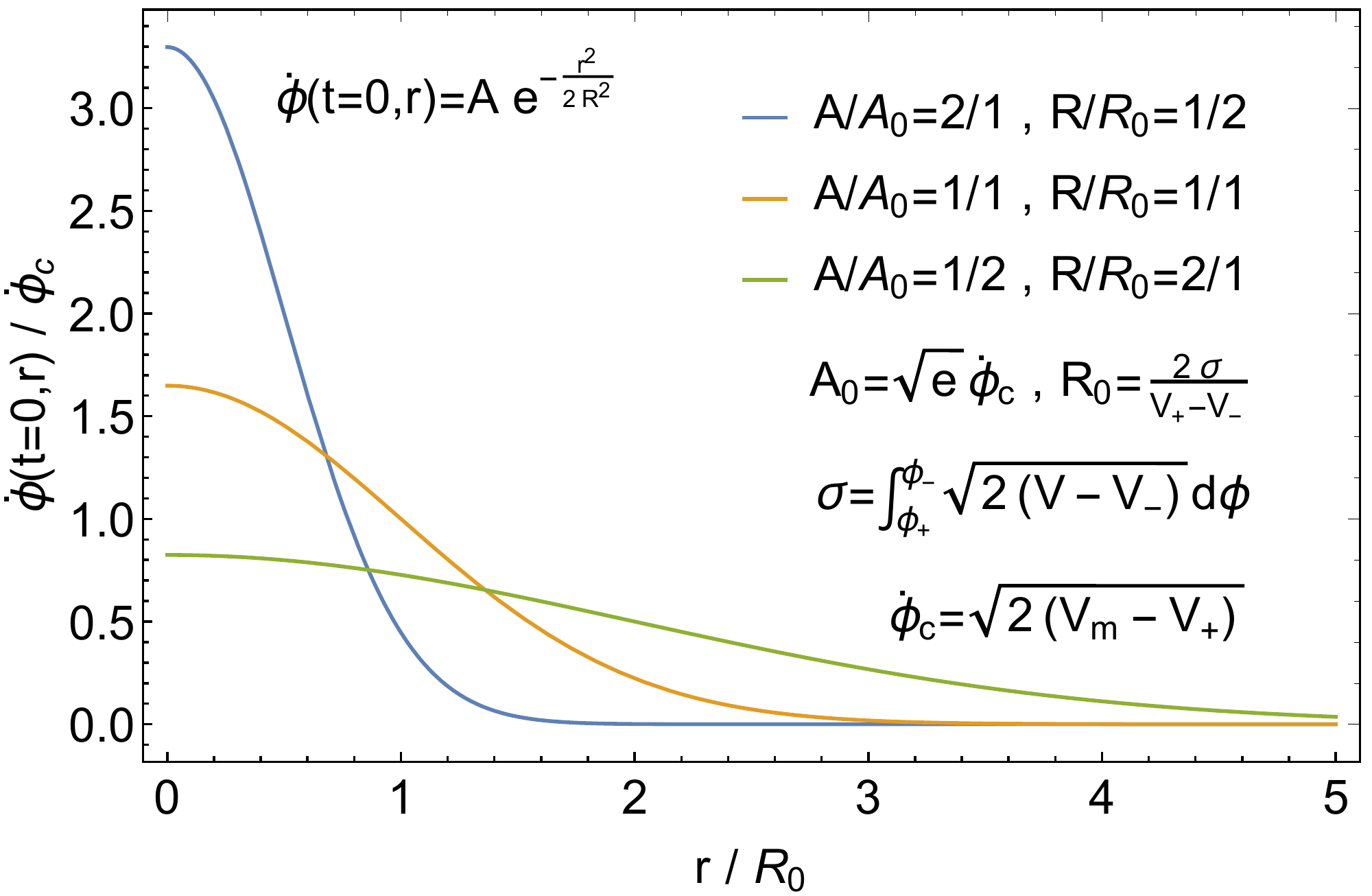}\\
\includegraphics[width=0.45\textwidth]{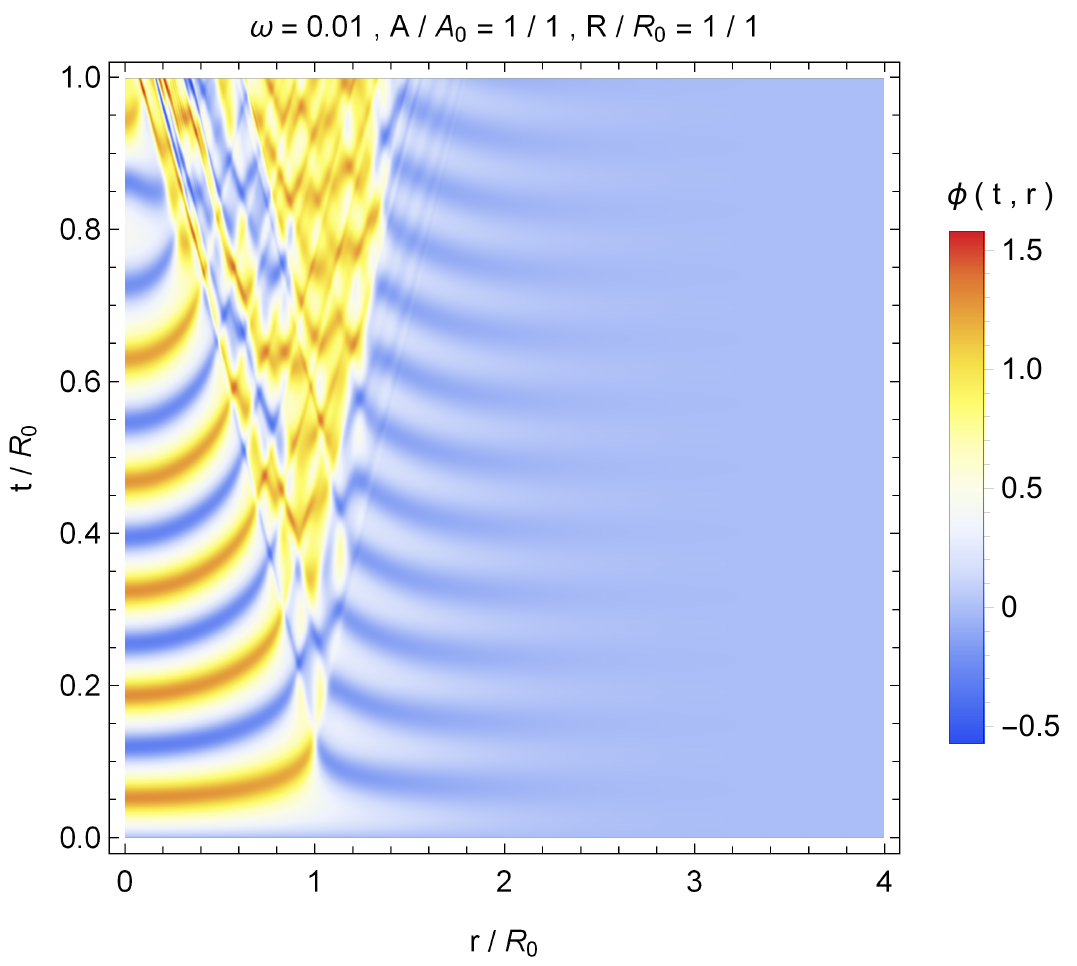}
\includegraphics[width=0.45\textwidth]{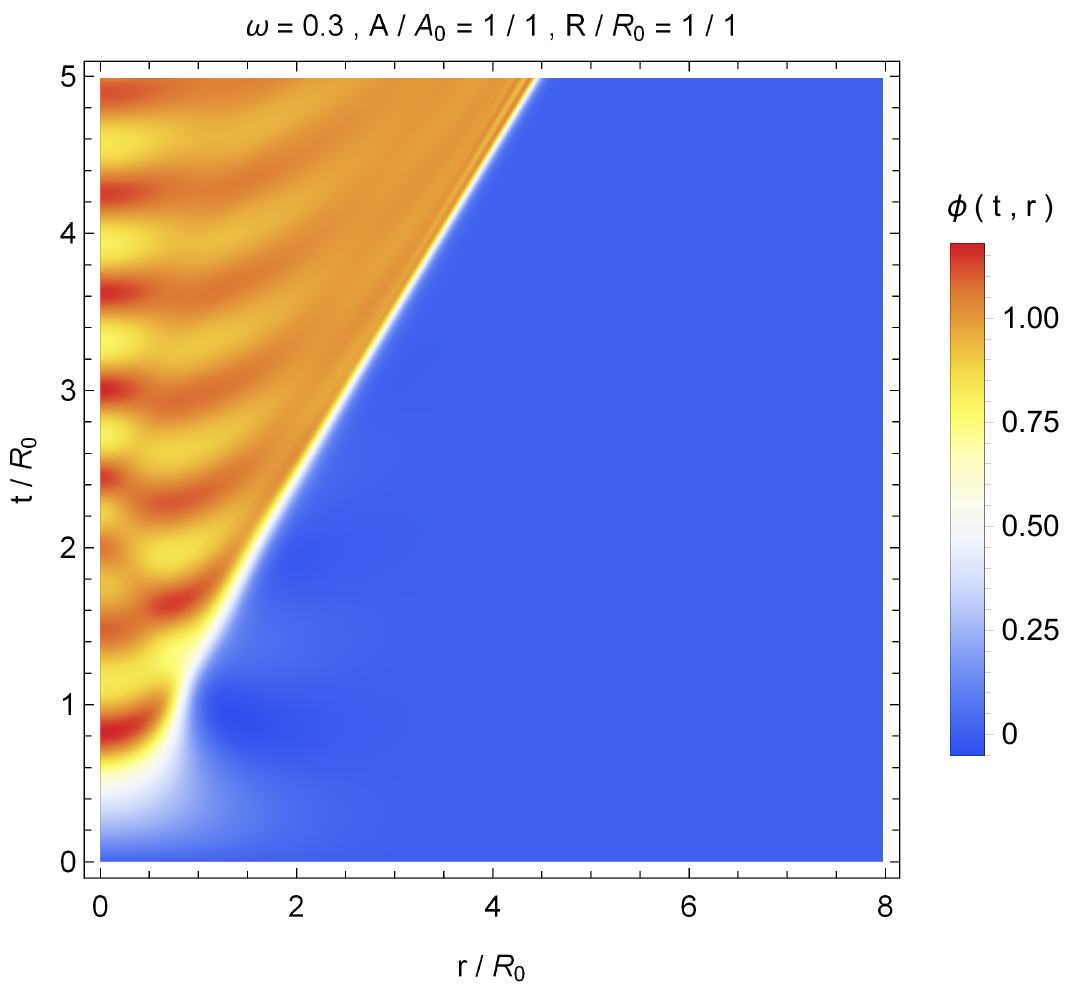}\\
\caption{Semiclassical picture of vacuum decay from an illustrative potential (upper left) with an initially vanished profile for the field value and Gaussian profile for the field velocity (upper right), whose classical evolution gives rise to the converted regions in true vacuum for both cases of thin wall (bottom left) and thick wall (bottom right).}\label{fig:thinthick}
\end{figure*}

To have a semiclassical vacuum decay, we first choose a test potential as
\begin{align}
V(\phi)=\phi^2-2(1+2\omega)\phi^3+(1+3\omega)\phi^4,
\end{align}
where the difference between the false vacuum $V(\phi_+)\equiv V_+=0$ at $\phi_+\equiv0$ and the true vacuum $V(\phi_-)\equiv V_-=-\omega$ at $\phi_-\equiv1$ is $\epsilon\equiv V_+-V_-=\omega$, while the height of potential barrier $V_b\equiv V_m-V_+$ is the difference between the false vacuum $V_+$ and the top of potential barrier $V_m=\frac{1}{16}(1+4\omega)/(1+3\omega)^3$ at $\phi_m=1/(2+6\omega)$. As shown in the upper left panel of Fig. \ref{fig:thinthick}, the thin-wall limit $V_b\gg\epsilon$ corresponds to $\omega\ll0.05$, while the thick-wall limit $V_b\ll\epsilon$ corresponds to $\omega\gg0.05$. We will take $\omega=0.01, 0.1, 0.3$ as illustrative examples for the cases of thin, intermediate, and thick walls. The wall thickness is conventionally estimated as $\delta\equiv\phi_m/\sqrt{V_b}=2\sqrt{(1+3\omega)/(1+4\omega)}$ to minimize the combination $[(\phi_m/\delta)^2+V_b]\delta$. 

The initial condition for flyover vacuum decay is assumed as
\begin{align}\label{eq:ini}
\phi(t=0,r)=0, \quad \dot{\phi}(t=0,r)=A\exp\left(-\frac{r^2}{2R^2}\right),
\end{align}
where the scalar field is initially sitting at the false vacuum homogeneously, and the field velocity (time derivative) occasionally develops from vacuum fluctuations a Gaussian profile with amplitude $A>\sqrt{2V_b}$ large enough to flyover the barrier directly due to $\frac12\dot{\phi}(t=0,r)^2>V_b$. For a bubble to expand, the width of the Gaussian profile $R>R_0$ is required to be larger than the minimal expansion radius $R_0=2\sigma/\epsilon$ with wall tension given by
\begin{align}
\sigma=\int_{\phi_+}^{\phi_-}\sqrt{2(V(\phi)-V_-)}\mathrm{d}\phi.
\end{align}

To uniformly specify the Gaussian profile of the initial field velocity, one could normalize it as shown in the upper right panel of Fig. \ref{fig:thinthick} by
\begin{align}
\dot{\phi}(t=0,r)=\left(\frac{A}{A_0}\right)A_0\exp\left[-\frac{(r/R_0)^2}{2(R/R_0)^2}\right],
\end{align}
where $A_0=\sqrt{2eV_b}$ is defined in such a way that the critical velocity $\dot{\phi}_c\equiv\sqrt{2V_b}$ is realized exactly at $r=R$ for $A=A_0$, as shown with the orange line in the upper right panel of Fig. \ref{fig:thinthick}. Therefore, for a general $A\neq A_0$, one could define a critical radius $r_c$, if at all, by the maximal $r$ within which the initial field velocity is larger than the critical velocity, namely, $\dot{\phi}(t=0, r\leq r_c)\geq\dot{\phi}_c$. It is easy to find that $r_c$ is given by
\begin{align}
r_c=R_0\sqrt{1+2\ln\frac{A}{A_0}}.
\end{align}
Hence, the initial Gaussian profile for field velocity could be nowhere larger than the critical velocity if $A/A_0<1/\sqrt{e}$, as shown with the green line in the upper right panel of Fig. \ref{fig:thinthick}. The blue line indicates a sub-$R_0$ region with supercritical velocity.

By numerically evolving the initial condition \eqref{eq:ini} with $A=A_0$ and $R=R_0$ via the classical equation of motion (EOM)
\begin{align}\label{eq:EOM}
\ddot{\phi}=\phi''+\frac{2}{r}\phi'-\frac{\mathrm{d}V}{\mathrm{d}\phi}
\end{align}
with boundary condition $\phi'(t,r=0)=0$, one obtains a stack of time slices of the field value as shown in the bottom panels of Fig. \ref{fig:thinthick} for the cases of thin (left) and thick (right) walls. For the thin-wall case, the bubble appears around $t\approx0.05R_0$ and then starts to expand around $t\approx0.1R_0$ with wall radius around $R_0$. The wall expands both inward and outward as observed in \cite{Blanco-Pillado:2019xny}. The inner transition region oscillates around the two minimums, whose radius eventually shrinks to zero around $t\approx R_0$. This picture is fairly different from the corresponding instanton case. However, for the thick-wall case, the picture is similar to the corresponding instanton case. The bubble appears around $t\approx0.6R_0$ and then starts to expand around $t\approx R_0$ with velocity approaching the speed of light. Remarkably, it is found in \cite{Blanco-Pillado:2019xny} that the probability to have an initial velocity profile larger than the critical velocity resembles the instanton estimation up to a numerical factor (although the numerical factor is larger in the thick-wall case than in the thin-wall case). However, in the following sections, we will take the thick-wall case as an example to present our finding, which should also apply to the thin-wall case, however, with more simulation cost.

\section{Semiclassical vacuum decay}\label{sec:semiclas}

\begin{figure*}
\centering
\includegraphics[width=0.53\textwidth]{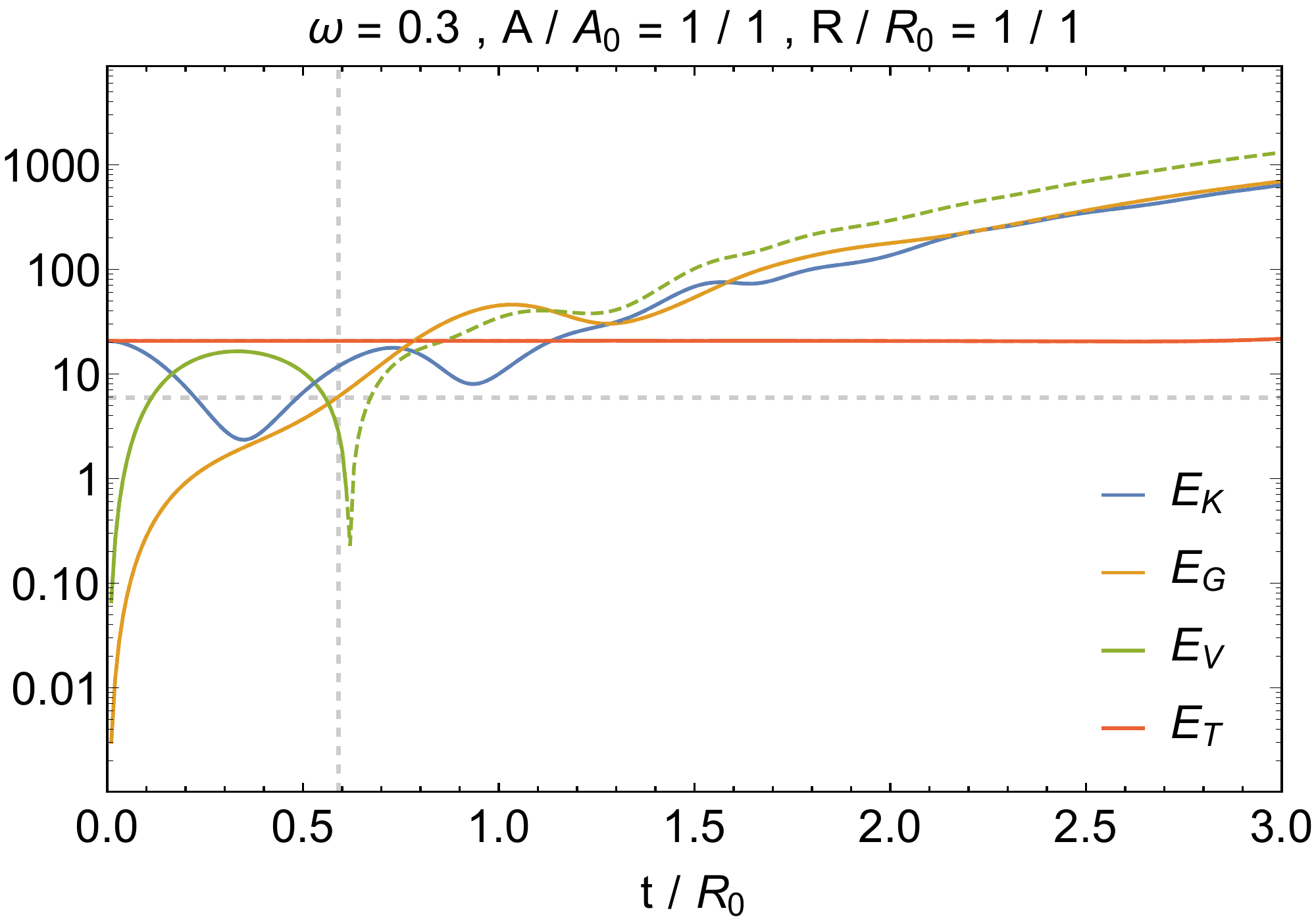}
\includegraphics[width=0.37\textwidth]{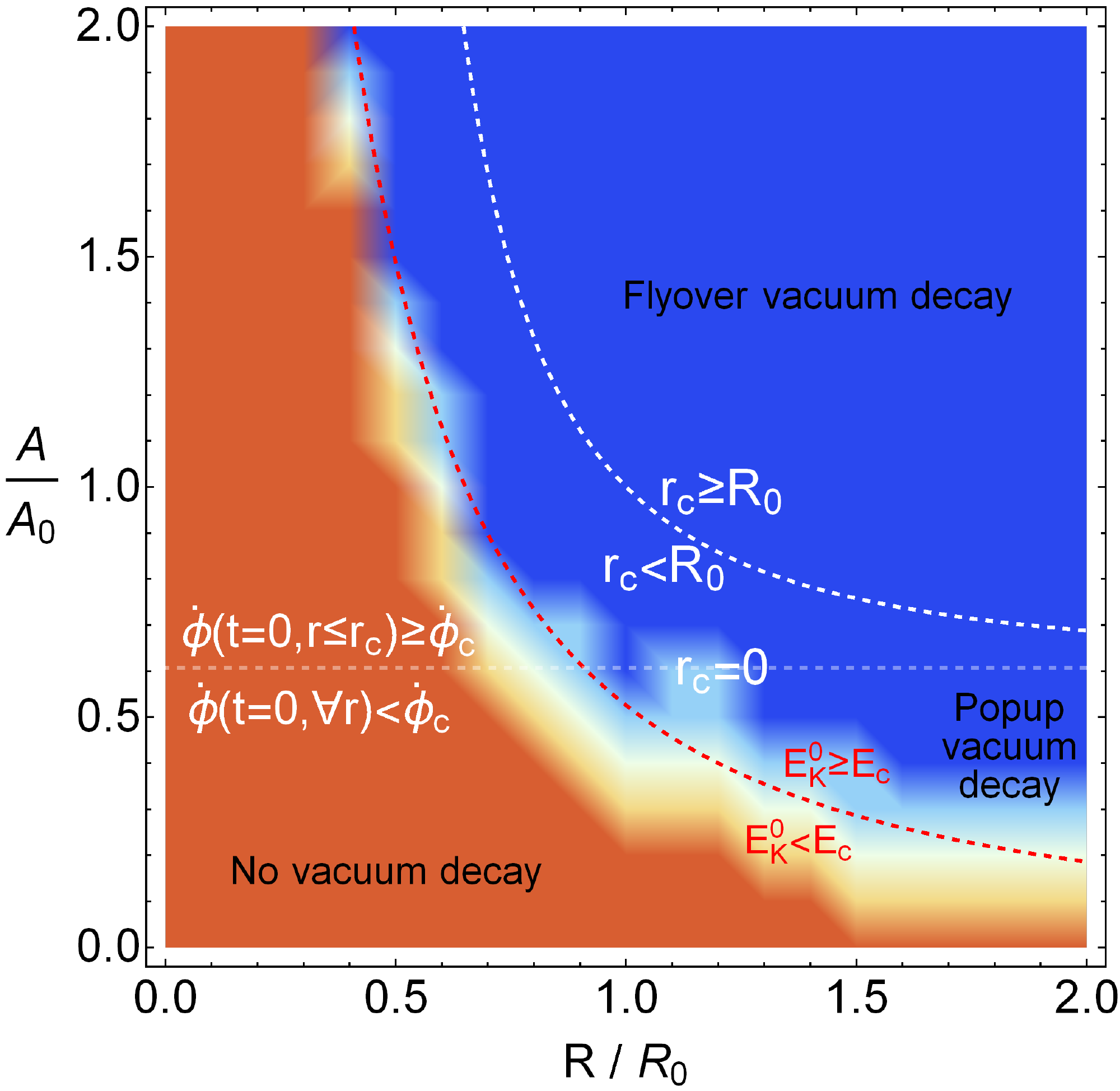}\\
\caption{\textit{Left}: the time evolutions of the integrated kinetic energy $E_K$, gradient energy $E_G$, potential energy $E_V$ (shown as dashed part of green curve when turning to negative) and total energy $E_T$ for the thick-wall case with $\omega=0.3$ and initial Gaussian profile of field velocity with $A=A_0$ and $R=R_0$. The vertical dashed line indicates the appearance time $t\approx0.6R_0$ of the transition region as shown in the bottom right panel of Fig. \ref{fig:thinthick}. The vertical dashed line intersects with $E_G$ at some critical value $E_c$ specified by the horizontal dashed line. \textit{Right}: the parameter space supported by the width $R/R_0$ and height $A/A_0$ of the initial Gaussian profile of field velocity. The white dashed horizontal line sets the boundary of vanishing critical radius $r_c$, below which there is nowhere in the initial velocity profile to be larger than the critical velocity required to classically overcome the barrier. The white dashed contour line signifies a critical radius $r_c=R_0$, above which is the originally proposed flyover vacuum decay. The red dashed contour line gives rise to a relaxed condition for the occurrence of semiclassical vacuum decay with the total kinetic energy larger than a critical energy, above which a pop-up vacuum decay could be possible even if it is below the white dashed horizontal line. We scan over the whole parameter space, where the orange region does not possess a semiclassical vacuum decay, while the blue region could have a transition region expanding outward. The fuzzy region is manifest due to a delayed time for bubble to appear that is exceeding the maximal simulation time limited by the numerical instability.}\label{fig:paraspace}
\end{figure*}

The flyover vacuum decay reviewed above is not limited to the critical case with $A=A_0$ and $R=R_0$ for the initial Gaussian profile of field velocity. It is suggested in \cite{Blanco-Pillado:2019xny} that the flyover vacuum decay occurs as long as the critical radius $r_c$, within which the initial field velocity is supercritical, is larger than the minimal expansion radius $R_0$. We will see that this only serves as a sufficient condition but not necessary. 

To qualify the necessary condition for the aforementioned realization of semiclassical vacuum decay, we present in the left panel of Fig. \ref{fig:paraspace} the full-time evolutions of the total kinetic energy $E_K$, gradient energy $E_G$, and potential energy $E_V$ over the whole simulation region of size $L$,
\begin{align}
E_K&=4\pi\int_0^L\frac12\dot{\phi}^2r^2\mathrm{d}r,\\
E_G&=4\pi\int_0^L\frac12\phi'^2r^2\mathrm{d}r,\\
E_V&=4\pi\int_0^LV(\phi)r^2\mathrm{d}r,
\end{align} 
where the conservative total energy $E_T=E_K+E_G+E_V$ is also computed as a consistency check for the numerical stability. It is easy to see that $E_T$ is initially equal to $E_K$ as it should be (apart from a compensation with a renormalized vacuum energy; see \cite{Hertzberg:2019wgx} for details), and $E_K$ is eventually converged with $E_G$ as $E_V$ becomes negative (dashed part of green curve). Since the transition region first appears around $t\approx0.6R_0$, as shown in the bottom right panel of Fig. \ref{fig:thinthick}, we plot a corresponding vertical dashed line, which intersects with $E_G$ at some value specified by a horizontal dashed line. We will denote this critical energy threshold as $E_c$, whose value will be estimated at the end of this section. The appearance of true vacuum bubble will be defined as the moment when $E_G$ eventually surpasses $E_c$.

A simple conjecture that could be imposed as the necessary condition expected for the semiclassical vacuum decay is that the initial total kinetic energy should be larger than some critical energy threshold $E_c$,
\begin{align}\label{eq:criteria1}
E_K^0=4\pi\int_0^\infty\frac12\dot{\phi}(t=0,r)^2r^2\mathrm{d}r=\frac{\pi^\frac32}{2}A^2R^3\geq E_c,
\end{align}
even if the kinetic energy density for each local degree of freedom is insufficient to classically flyover the barrier. The key factor is the nonlinear evolution of EOM \eqref{eq:EOM} that converts some of the kinetic energy into the gradient energy to eventually form a bubble. The central task is to determine the precise form/value of $E_c$. 

One of the approaches to determine the existence of $E_c$ is to scan over the whole parameter space of the initial Gaussian profile of field velocity parametrized by $R/R_0$ and $A/A_0$, as shown in the right panel of Fig. \ref{fig:paraspace}. A vanishing $r_c$ is specified by the white dashed horizontal line with $A/A_0=1/\sqrt{e}$, below which the whole initial profile of field velocity is smaller than the critical velocity. A comparable $r_c$ to the minimal expansion radius $R_0$ is specified by the white dashed contour line, above which is the originally proposed flyover vacuum decay. However, as specifically checked by numerical simulations with all different combination values of $R/R_0$ and $A/A_0$, the blue region is successful to implement the semiclassical vacuum decay, which is obviously larger than the original region suggested by $r_c\geq R_0$. The semiclassical vacuum decay is possible for $r_c$ smaller than the minimal expansion radius $R_0$ and still possible even when $r_c$ does not exist. The orange region does not possess a semiclassical vacuum decay, which is separated from the blue region by a fuzzy zone. However, the fuzzy zone is manifest itself only for a practical reason that the time for the bubble to appear, if at all, has exceeded the maximal simulation time limited by the numerical instability. Nevertheless, since there is either an expanding bubble or not, this fuzzy zone should be theoretically a sharp transition line as shown with the red dashed contour line by $E_K^0=E_c$. 

\subsection{Flyover vacuum decay}\label{subsec:flyover}

\begin{figure*}
\centering
\includegraphics[width=0.4\textwidth]{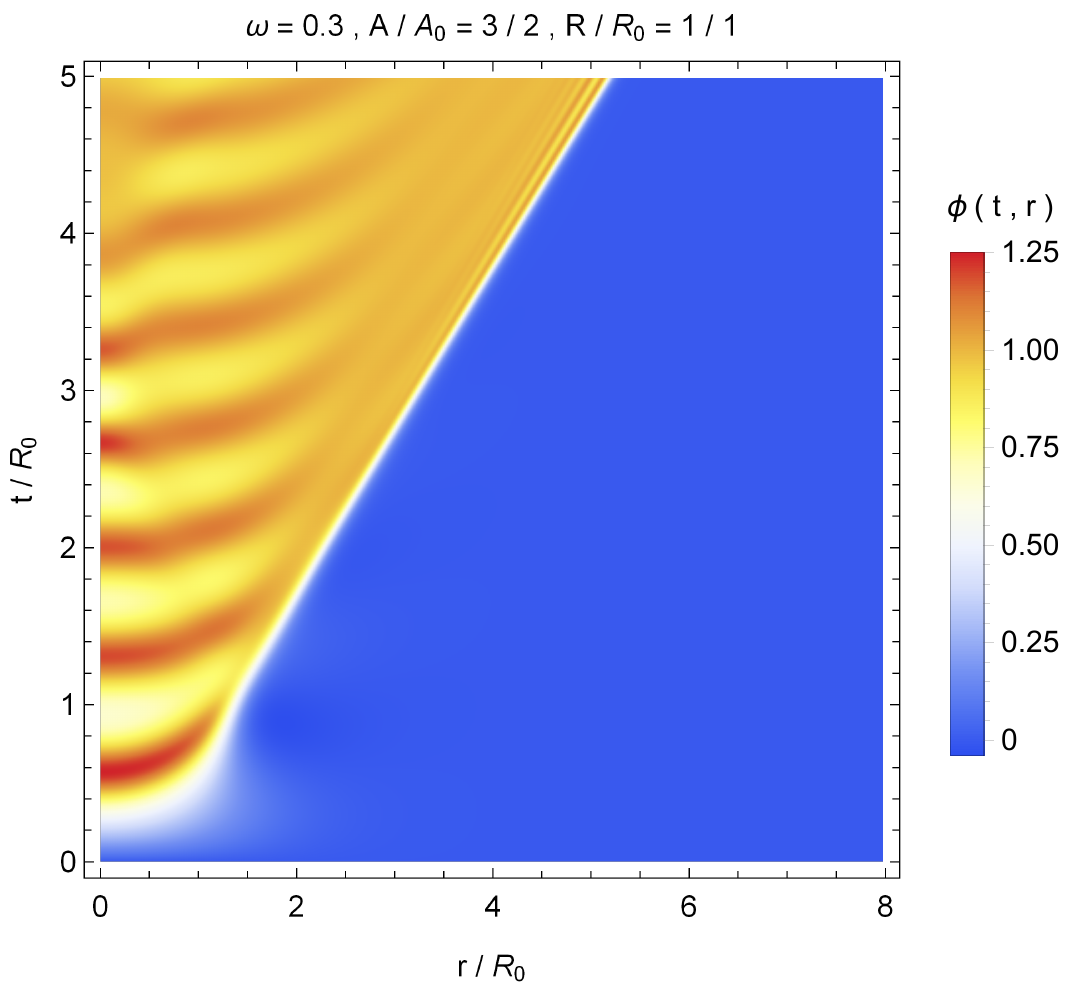}
\includegraphics[width=0.5\textwidth]{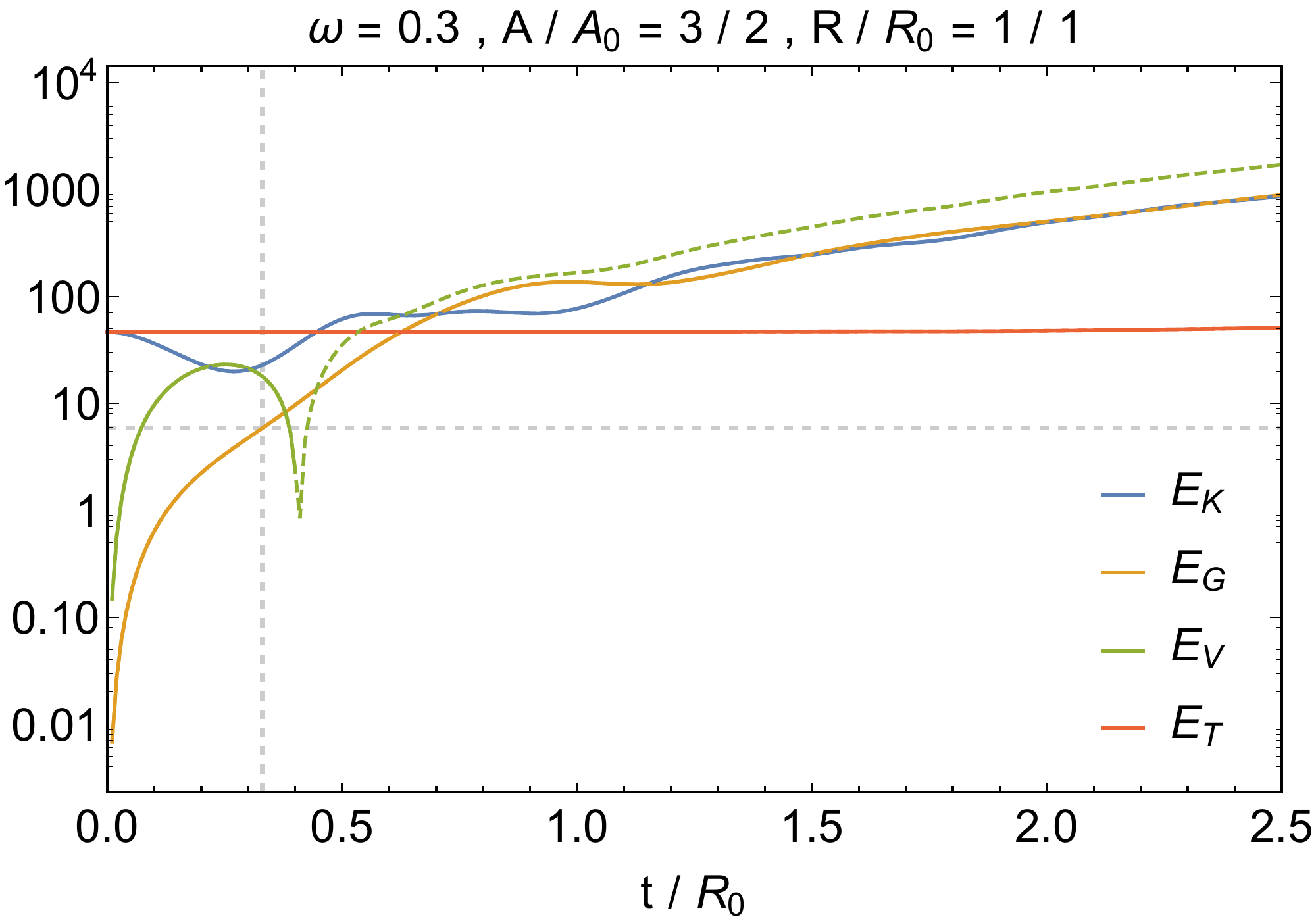}\\
\includegraphics[width=0.4\textwidth]{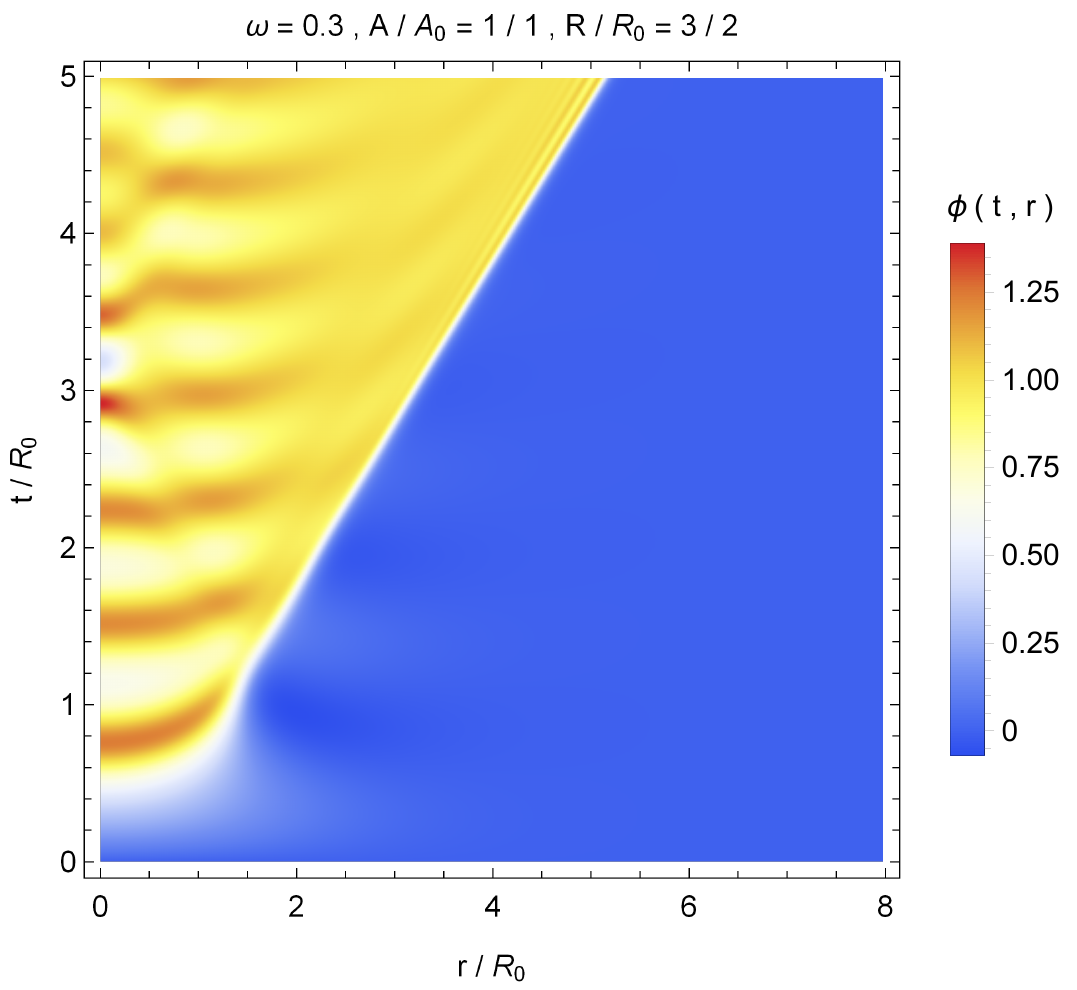}
\includegraphics[width=0.5\textwidth]{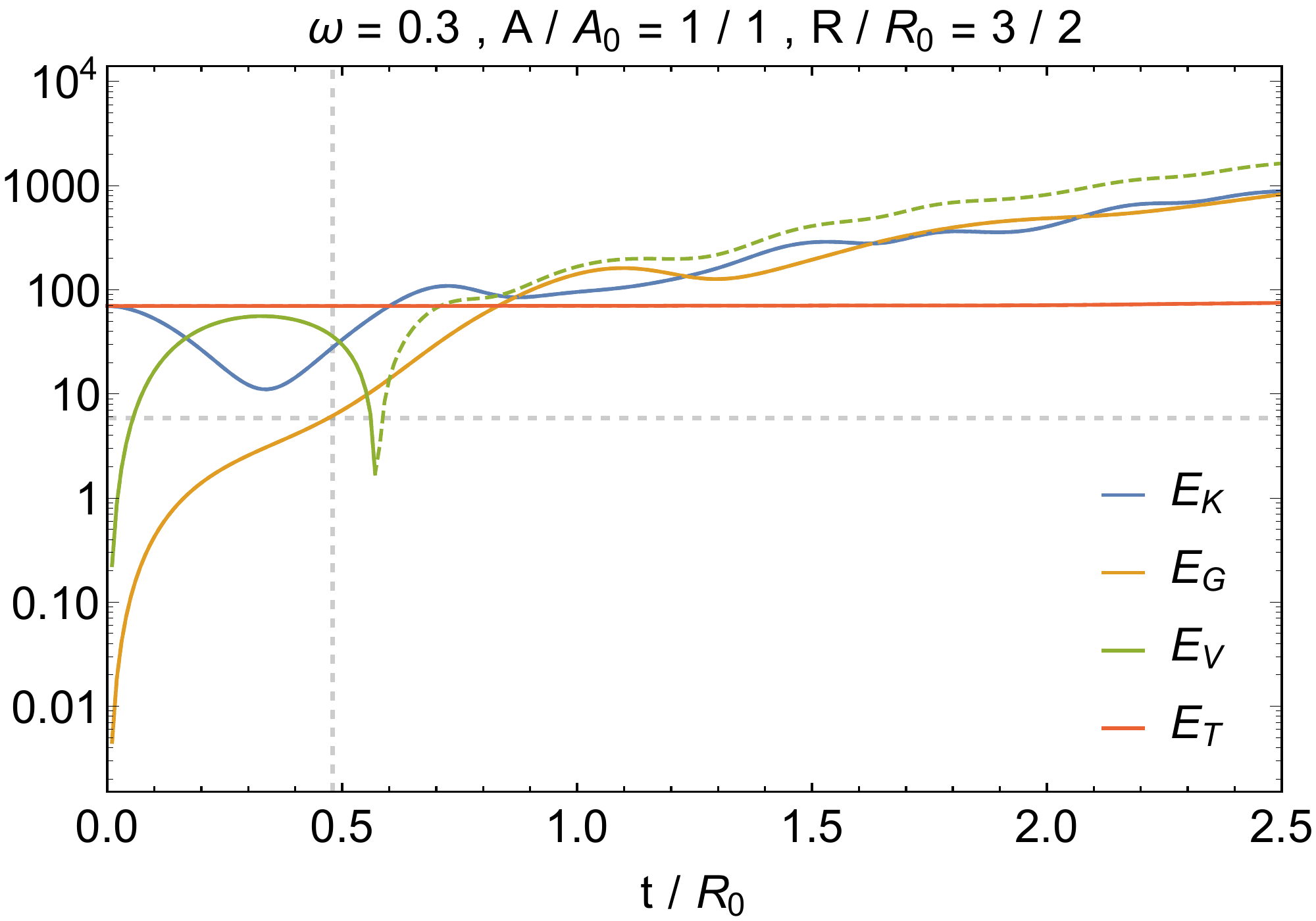}\\
\includegraphics[width=0.4\textwidth]{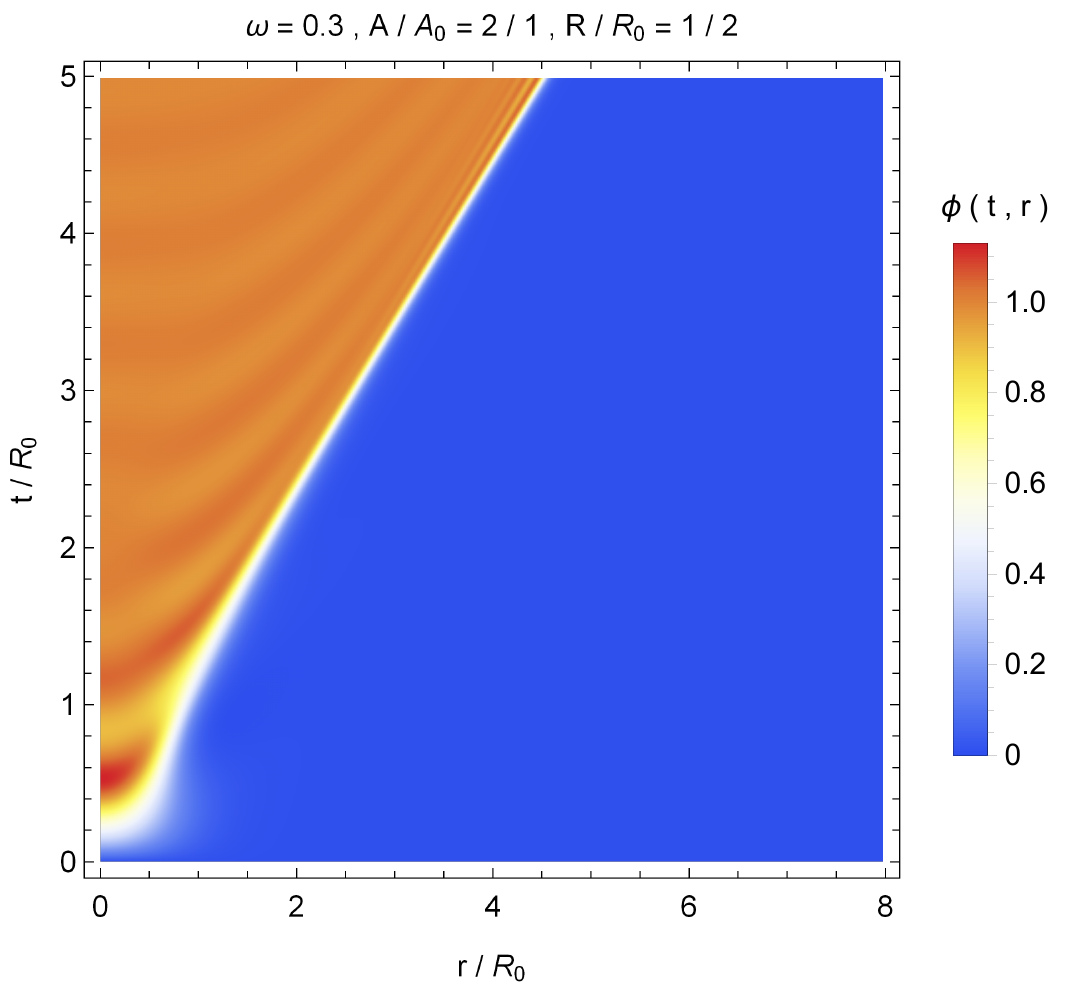}
\includegraphics[width=0.5\textwidth]{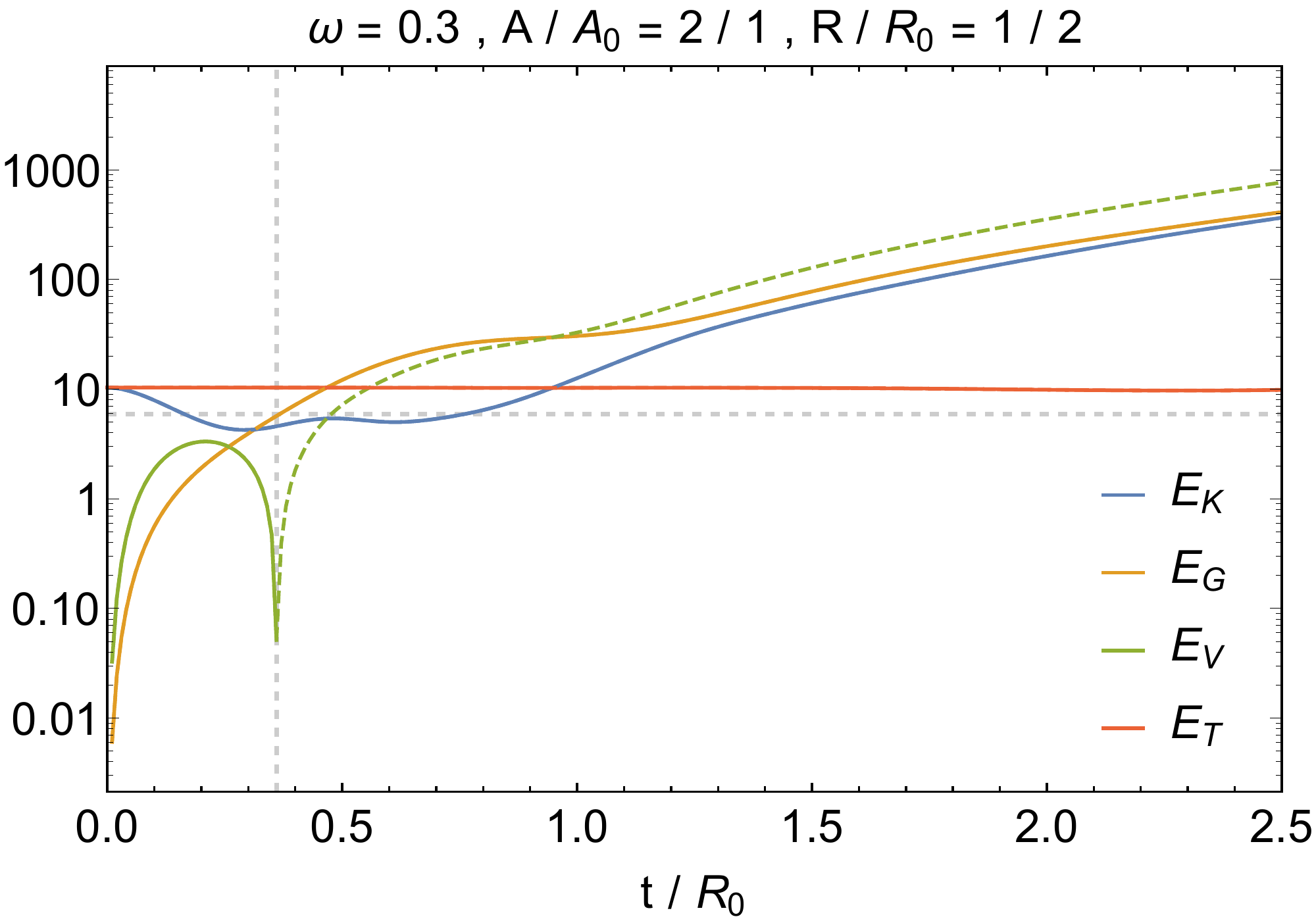}\\
\caption{Three illustrative examples of general flyover vacuum decay with $r_c\geq0$ and $E_K^0\geq E_c$, where the time evolutions of field value (left column) and integrated energies (right column) are presented. The first two examples have $r_c>R_0$, while the third one has $r_c<R_0$ by our parameter choices specified in the figures.  In the right column, the dashed part of green curve is the opposite value of potential energy, the vertical dashed line is the bubble appearance time, and the horizontal dashed line is the critical energy threshold $E_c$.}\label{fig:flyover}
\end{figure*}

The region enclosed by $r_c\geq0$ and $E_K^0\geq E_c$ could be regarded as a general type of flyover vacuum decay, since there is a region of size $r_c$ where field velocity is supercritical, which is large enough for any local degree of freedom inside $r_c$ to classically flyover the potential barrier directly. Three examples of such general type of flyover vacuum decay are given in Fig. \ref{fig:flyover}, the first two of which are the original type of flyover vacuum decay, while the third one has $r_c$ below $R_0$. The difference lies in the bubble radius just before the expansion at the speed of light, which is larger than $R_0$ in the first two examples while smaller than $R_0$ in the third example. Aside from the time evolutions of field value in the left column, we also present in the right column the time evolutions of the integrated kinetic energy, gradient energy, potential energy, and total energy. The dashed part of green curve denotes the absolute value of a negative potential energy when the bubble expands after its appearance. The kinetic energy is eventually converged with the gradient energy after long-time expansion. The horizontal dashed line is the same critical energy threshold $E_c$, as shown in Fig. \ref{fig:paraspace}, which always intersects $E_G$ exactly at the time when the bubble appears. This will be a pattern that keeps showing up in the numerical simulations, which also reinforces the very existence of such a critical energy threshold $E_c$.

\subsection{Pop-up vacuum decay}\label{subsec:popup}

\begin{figure*}
\centering
\includegraphics[width=0.4\textwidth]{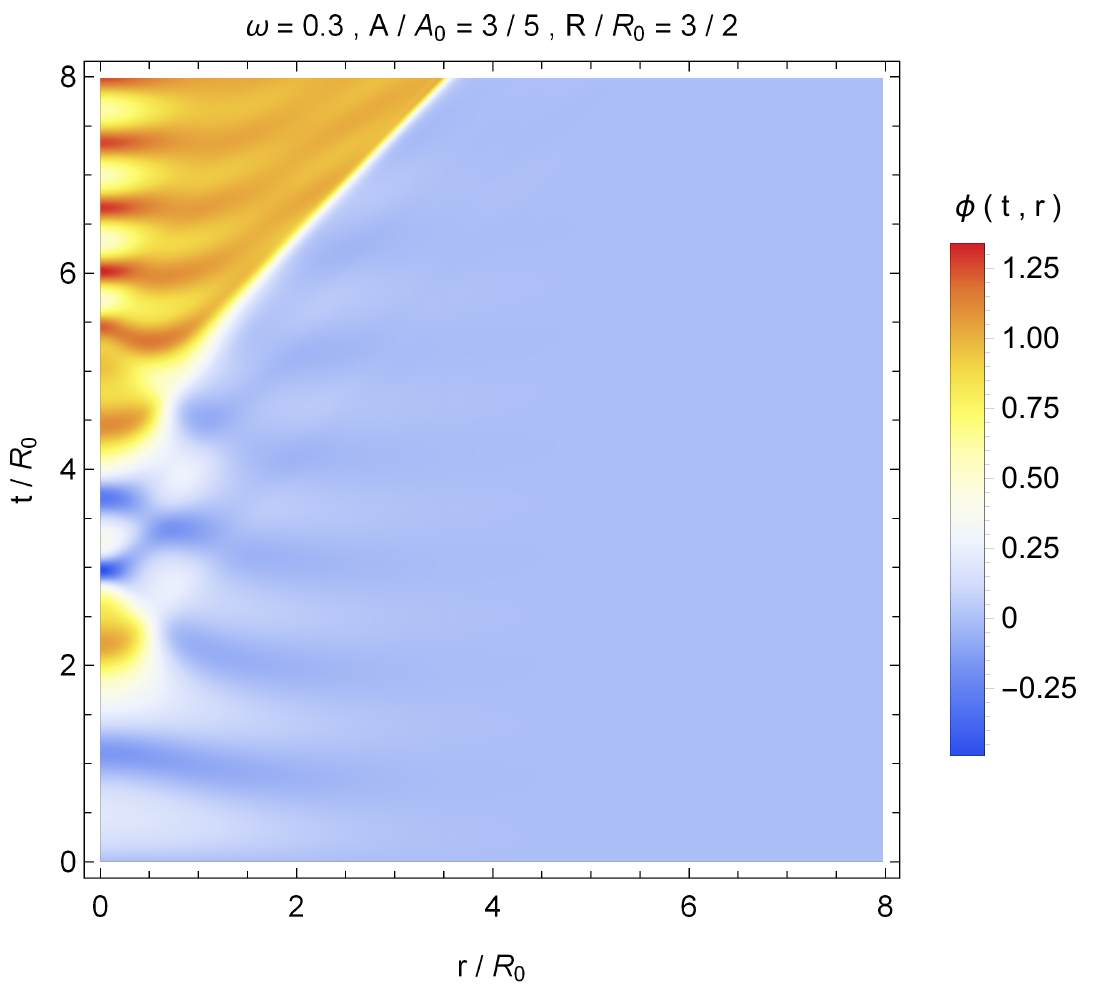}
\includegraphics[width=0.5\textwidth]{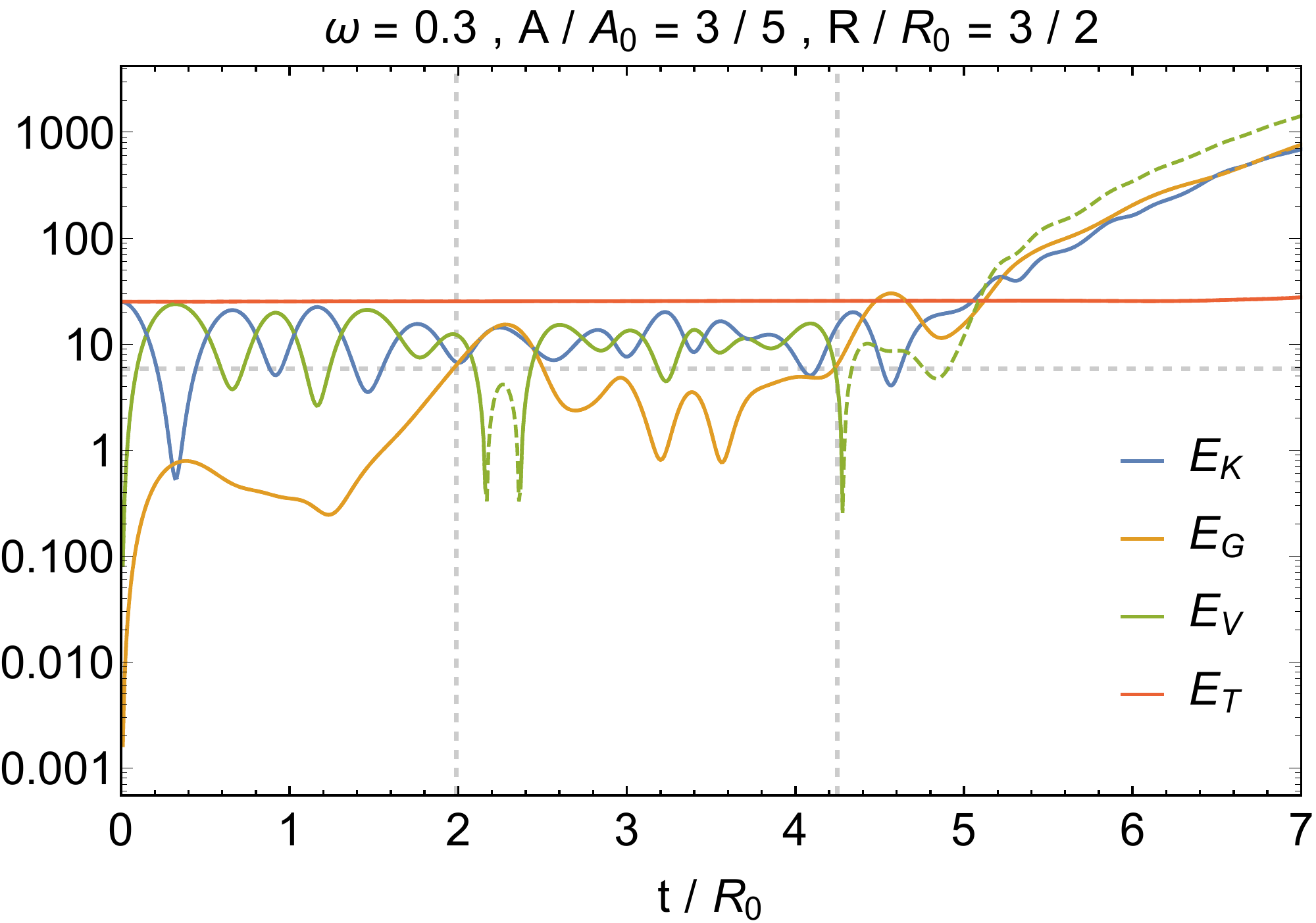}\\
\includegraphics[width=0.4\textwidth]{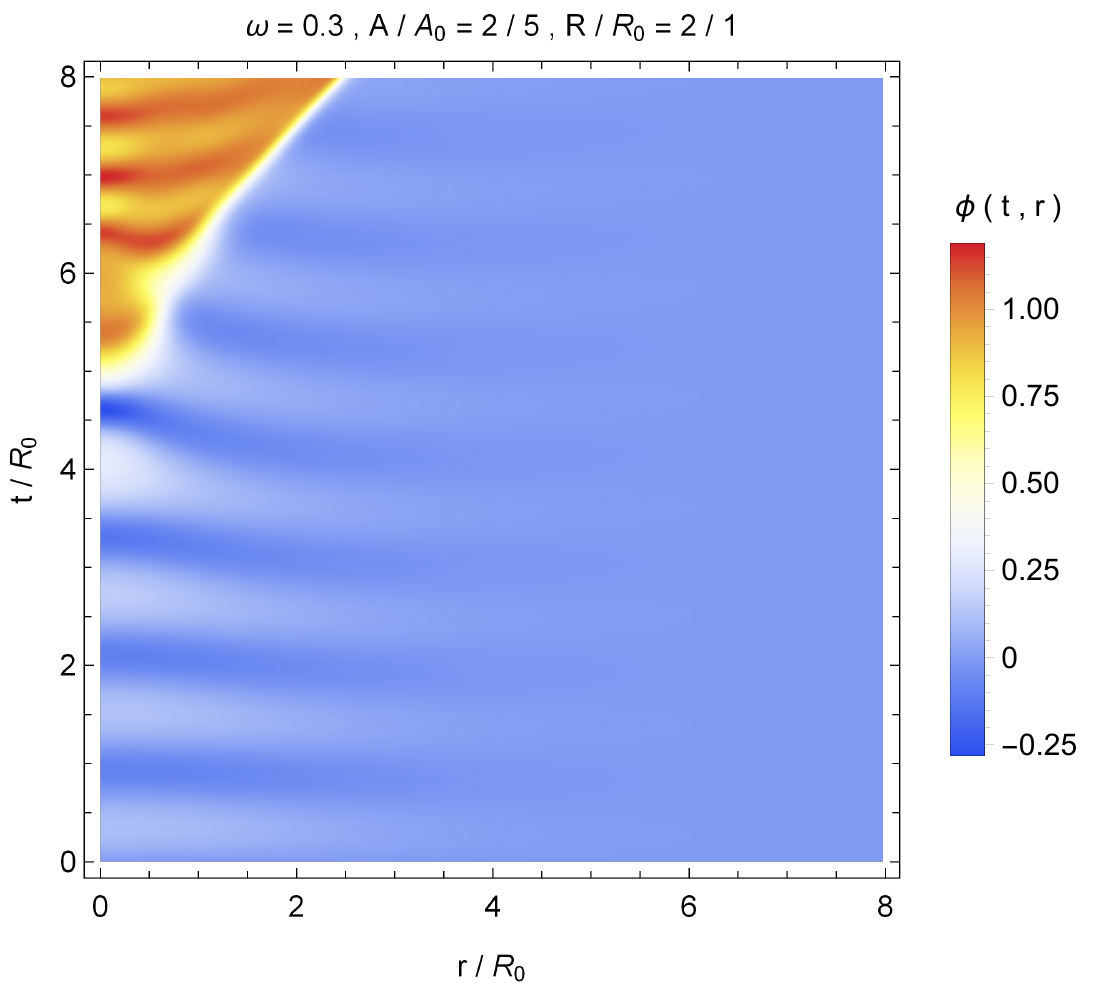}
\includegraphics[width=0.5\textwidth]{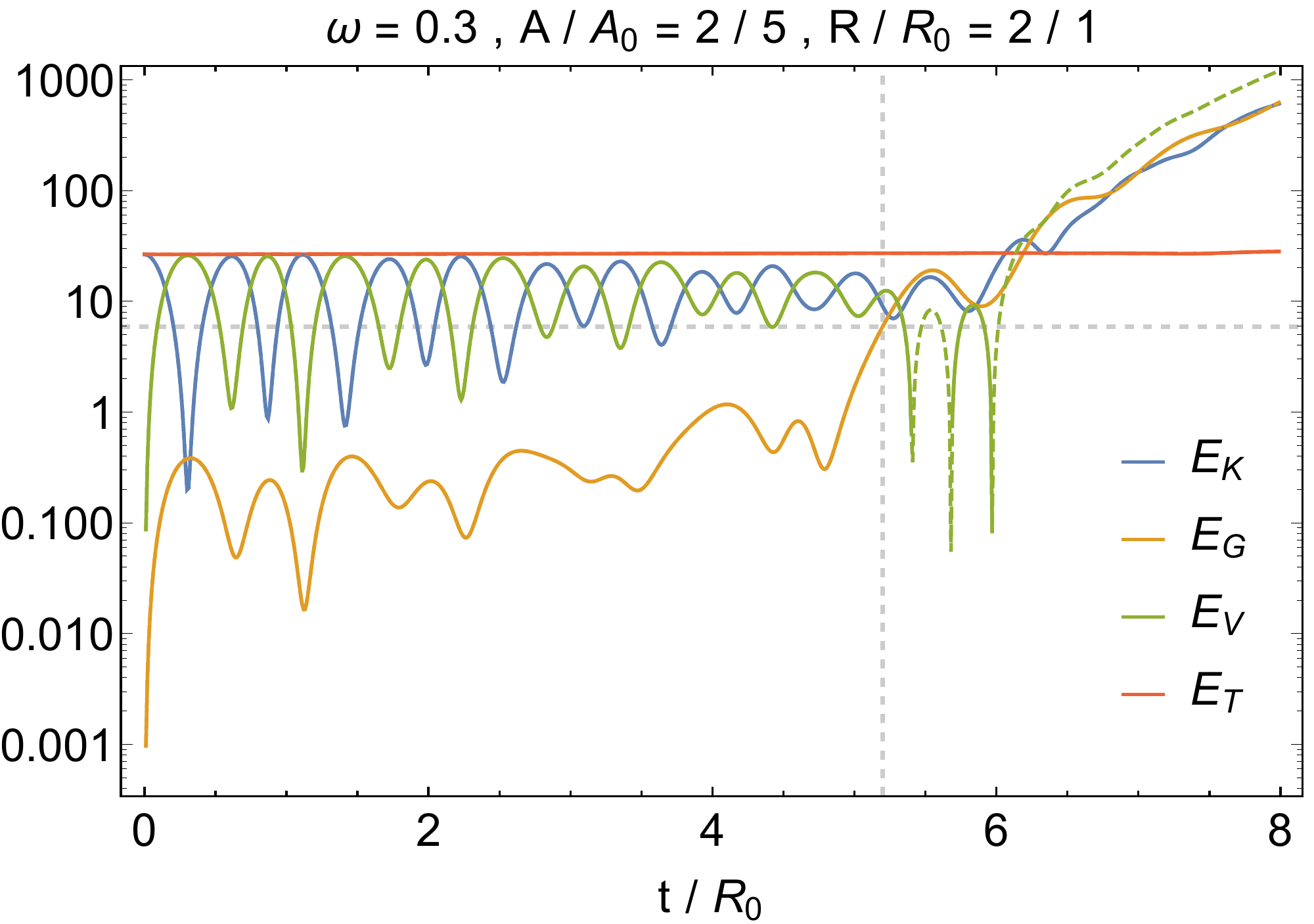}\\
\includegraphics[width=0.4\textwidth]{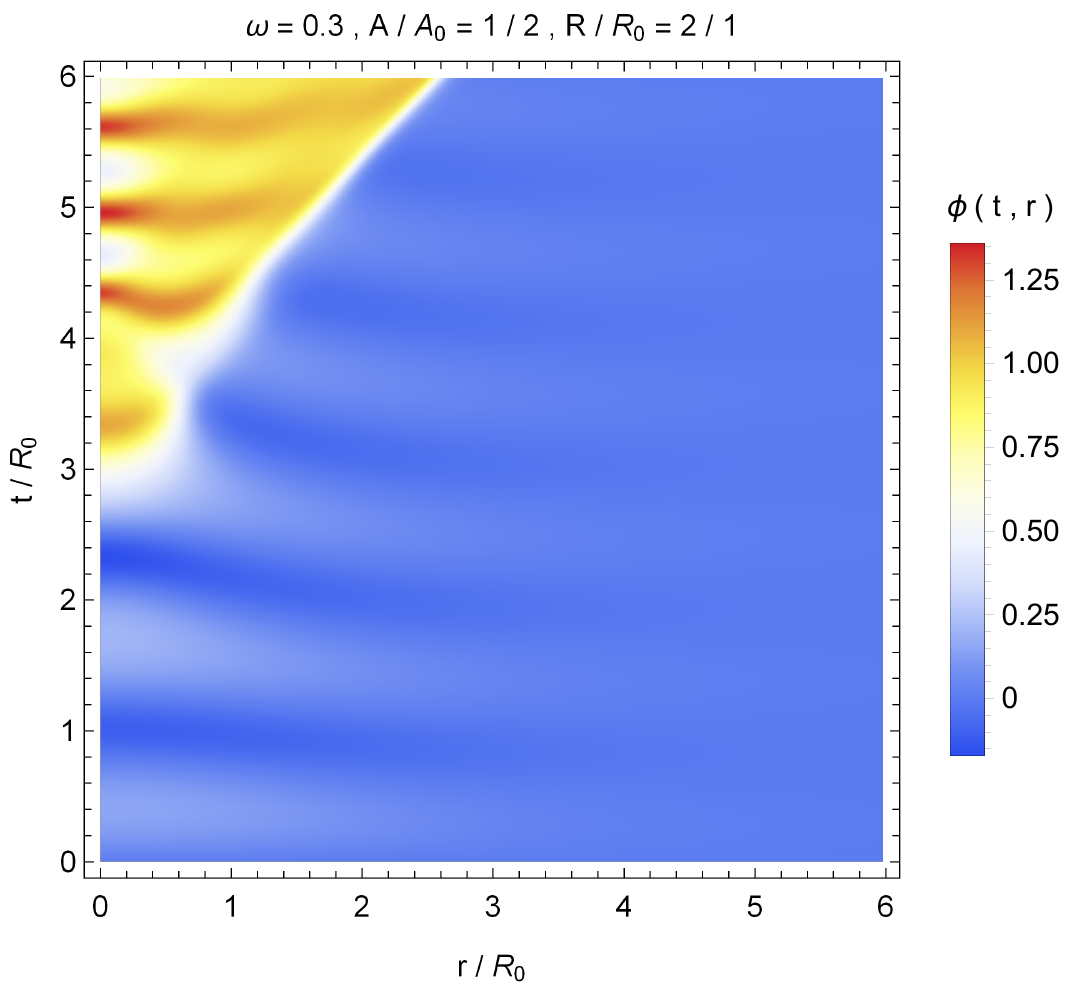}
\includegraphics[width=0.5\textwidth]{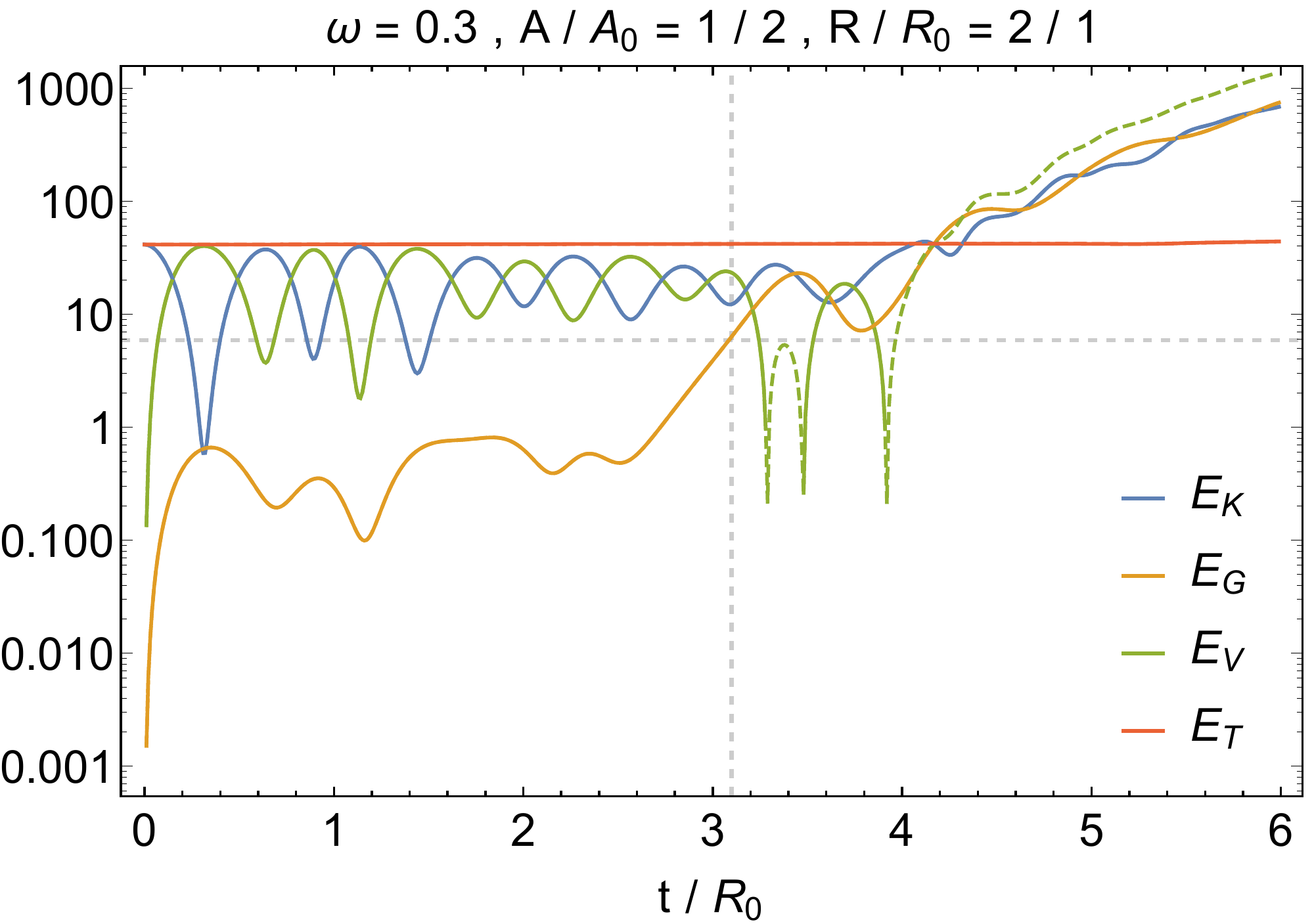}\\
\caption{Three illustrative examples of pop-up vacuum decay with $E_K^0\geq E_c$ but without $r_c$ (namely, the initial profile of field velocity is nowhere larger than the critical velocity), where the time evolutions of field value (left column) and the integrated energies (right column) are presented. In the right column, the dashed part of green curve is the opposite value of potential energy, the vertical dashed line is the bubble appearance time, and the horizontal dashed line is the critical energy threshold $E_c$.}\label{fig:popup}
\end{figure*}

\begin{figure*}
\centering
\includegraphics[width=0.45\textwidth]{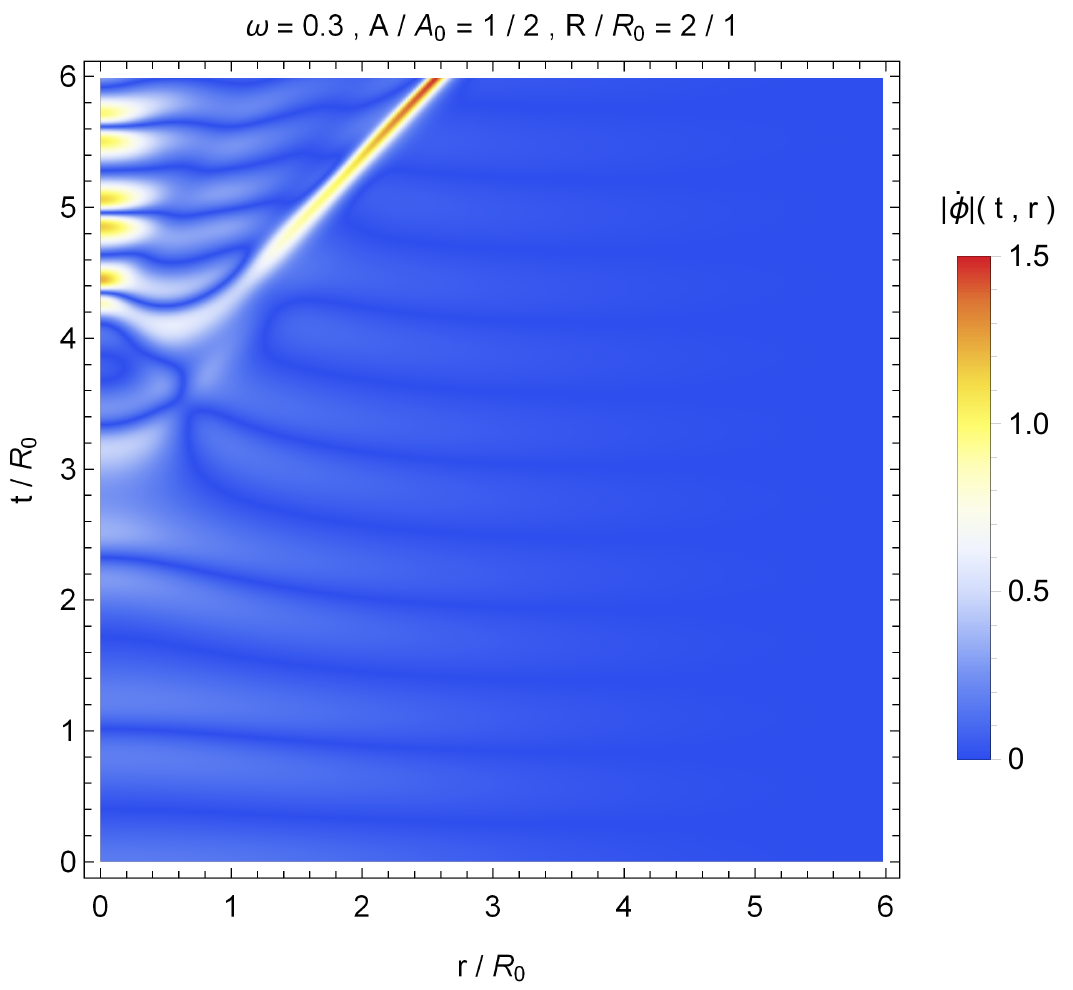}
\includegraphics[width=0.45\textwidth]{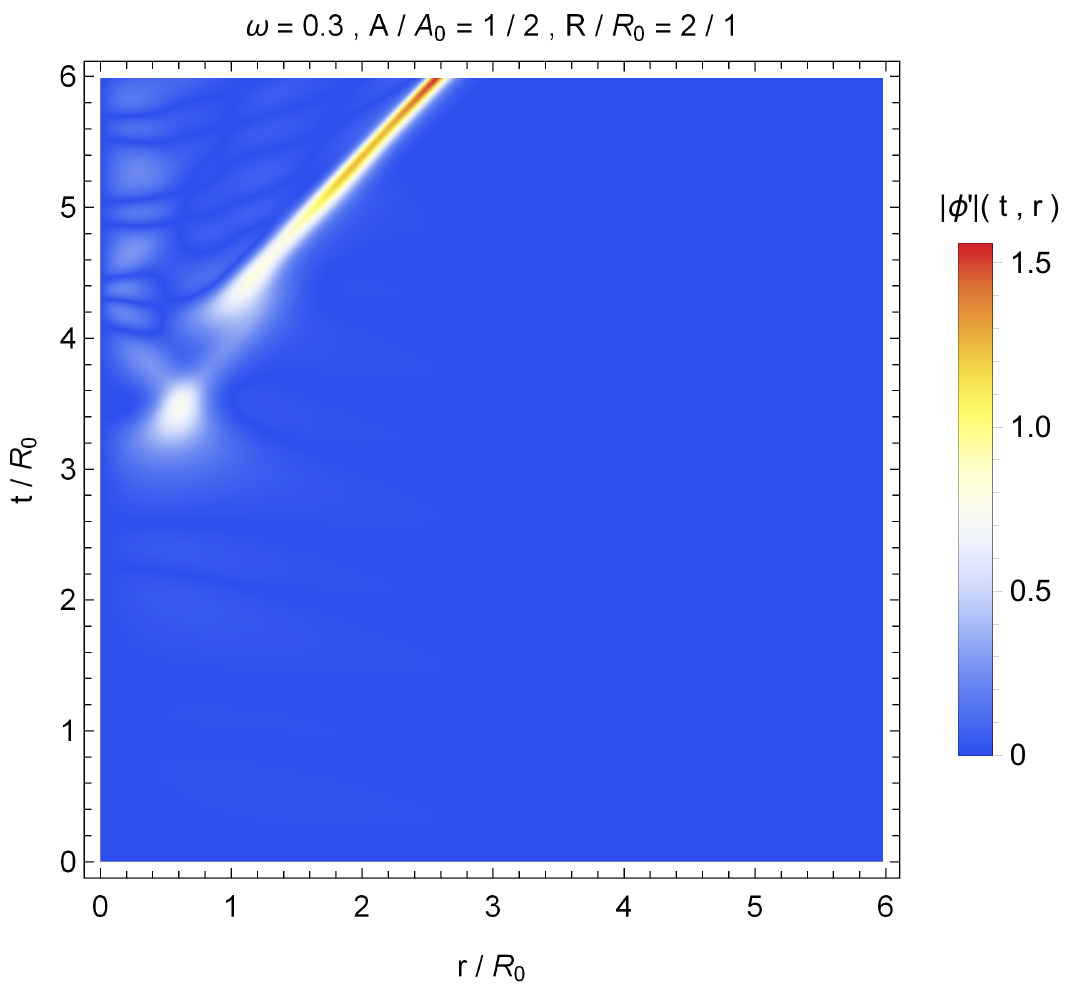}\\
\caption{The time evolutions of kinetic term $|\dot{\phi}|$ (left) and gradient term $|\phi'|$ (right) of the pop-up vacuum decay shown in the third example of Fig. \ref{fig:popup}.}\label{fig:popup2}
\end{figure*}

For the region enclosed by $E_K^0\geq E_c$ but below $r_c=0$, we also find the appearance of bubble with subsequent expansion, even if the initial profile of field velocity is everywhere insufficient to classically overcome the potential barrier directly. We also present three examples of such semiclassical vacuum decay in Fig. \ref{fig:popup}, where the time evolutions of field value (left column) and the integrated energies (right column) are presented with the same convention as in the previous sections. In all three examples, compared to the flyover cases, the appearance of bubble is significantly delayed, before which the total kinetic energy and potential energy are oscillating with decreasing amplitudes that are likely converted into the gradient energy.  Finally, the bubble appears when the gradient energy exceeds the same critical energy threshold $E_c$ presumed in the previous sections. Soon after the appearance of bubble, the total potential energy quickly turns to zero before becoming negative as bubble expanding. An intriguing case is the first example, where the bubble appears and disappears around $t\approx2R_0$ and then reappears around $t\approx4R_0$ with successful expansion followed. It seems that the expanding bubble finally appears after a (or several) short pop-up(s) of collapsing bubble(s), hence the name of \textit{pop-up vacuum decay}.

To see how such a bubble could appear even if the initial kinetic energy density is everywhere insufficient to classically overcome the barrier, the time evolutions of the to-be-integrated kinetic term $|\dot{\phi}|$ and gradient term $|\phi'|$ at each time slice are presented in Fig. \ref{fig:popup2} for the last example of Fig. \ref{fig:popup}. Near the emergence of bubble wall, the kinetic term $|\dot{\phi}|$ looks like some dark blue valleys surrounded by the white ridges, while the gradient term $|\phi'|$ is suddenly enlarged as a white spark around $t\approx3.2R_0$ after gradually gathering itself around the would-be wall region through the decreasing kinetic energy. Therefore, the initial total kinetic energy $E_K^0=\frac12\pi^{3/2}A^2R^3$ should be at least larger than the total gradient energy at the onset of wall appearance as the critical energy threshold $E_c$,
\begin{align}
E_G=4\pi\int\frac12\phi'^2r^2\mathrm{d}r\sim\left(\frac{\phi_m}{\delta}\right)^2R_0^2\delta\sim V_b R_0^2\delta\equiv E_c.
\end{align}
The criterion \eqref{eq:criteria1} now gives rise to a form of
\begin{align}\label{eq:criteria2}
\left(\frac{A}{A_0}\right)^2\left(\frac{R}{R_0}\right)^3\gtrsim\frac{\delta}{R_0}
\end{align}
up to some universal numerical factor. In particular, for the thick-wall case, we have tested so far with comparable thickness to bubble radius $\delta\sim R_0$, the total gradient energy as the critical energy threshold $E_c$ could be rewritten as
\begin{align}
E_G\sim V_b R_0^2\delta\sim\frac{4\pi}{3}V_bR_0^3\equiv E_c,
\end{align}
which is the total potential energy with converted region of minimal expansion size at the top of potential barrier. The criterion \eqref{eq:criteria2} is now replaced by
\begin{align}\label{eq:criteria3}
\left(\frac{A}{A_0}\right)^2\left(\frac{R}{R_0}\right)^3\gtrsim\frac{4}{3e\sqrt{\pi}},
\end{align}
which is shown as the red dashed curve in the right panel of Fig. \ref{fig:paraspace}.
It is worth noting that, our criteria for the appearance of an expanding bubble are expressed in terms of a combination $\sim A^2R^3$, while the flyover transition rate found in \cite{Blanco-Pillado:2019xny} goes as $\sim \exp(-A^2 R^3)$. This suggests that transitions with different $A$ and $R$ but the same value of $A^2 R^3$ may have nearly the same probability shown in the second panel of Fig.\ref{fig:paraspace}.

\subsection{No semiclassical vacuum decay}\label{subsec:nodecay}

\begin{figure*}
\centering
\includegraphics[width=0.4\textwidth]{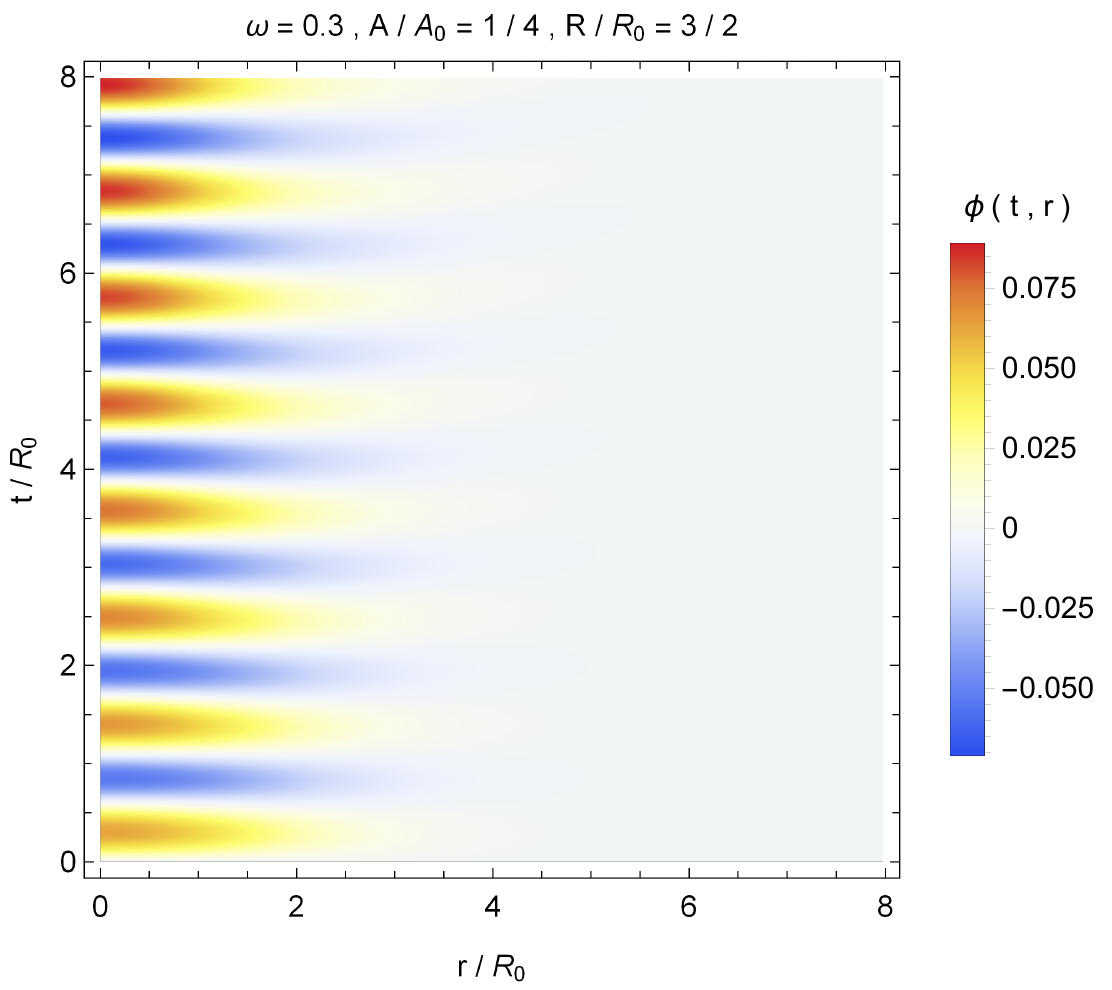}
\includegraphics[width=0.5\textwidth]{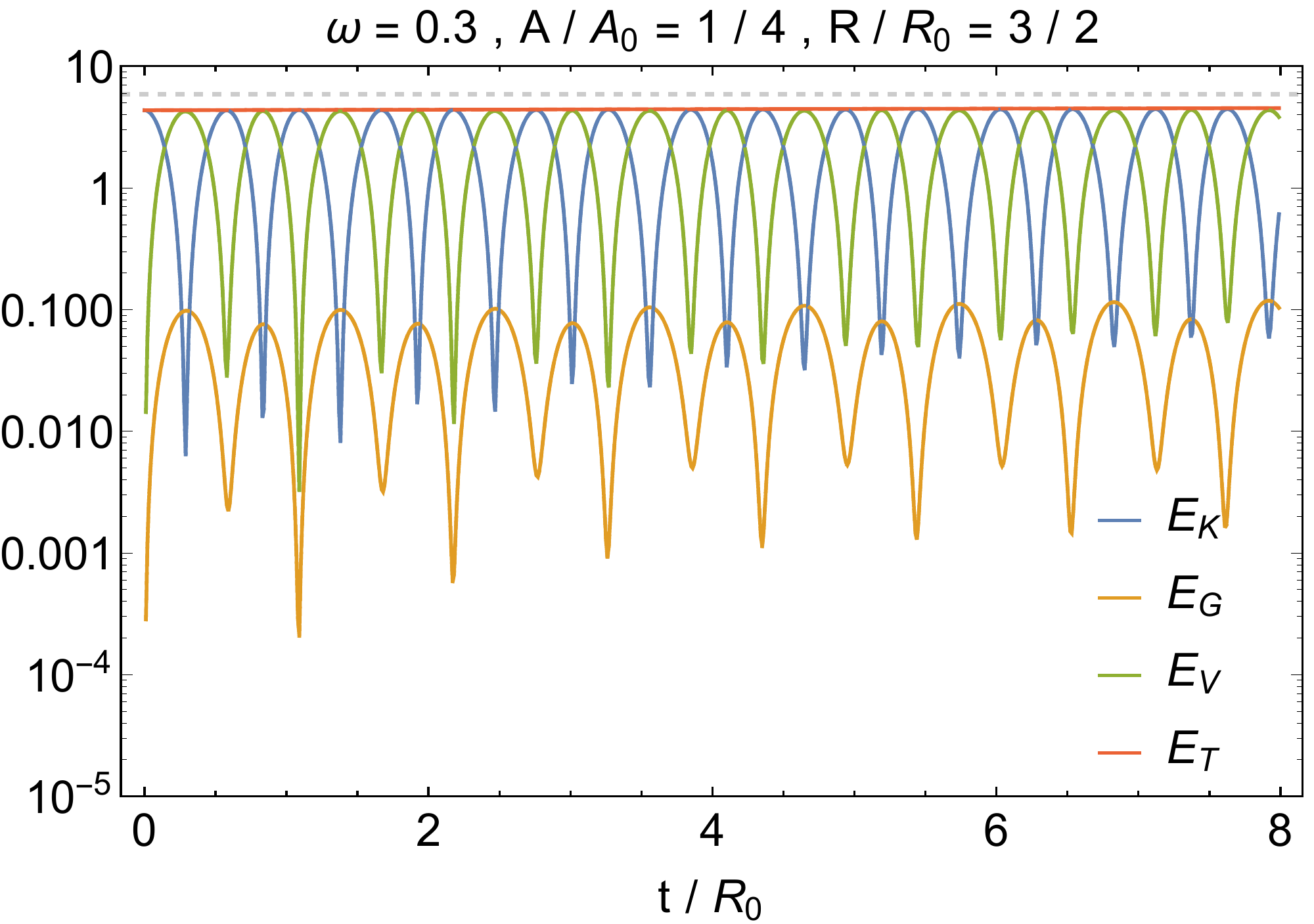}\\
\includegraphics[width=0.4\textwidth]{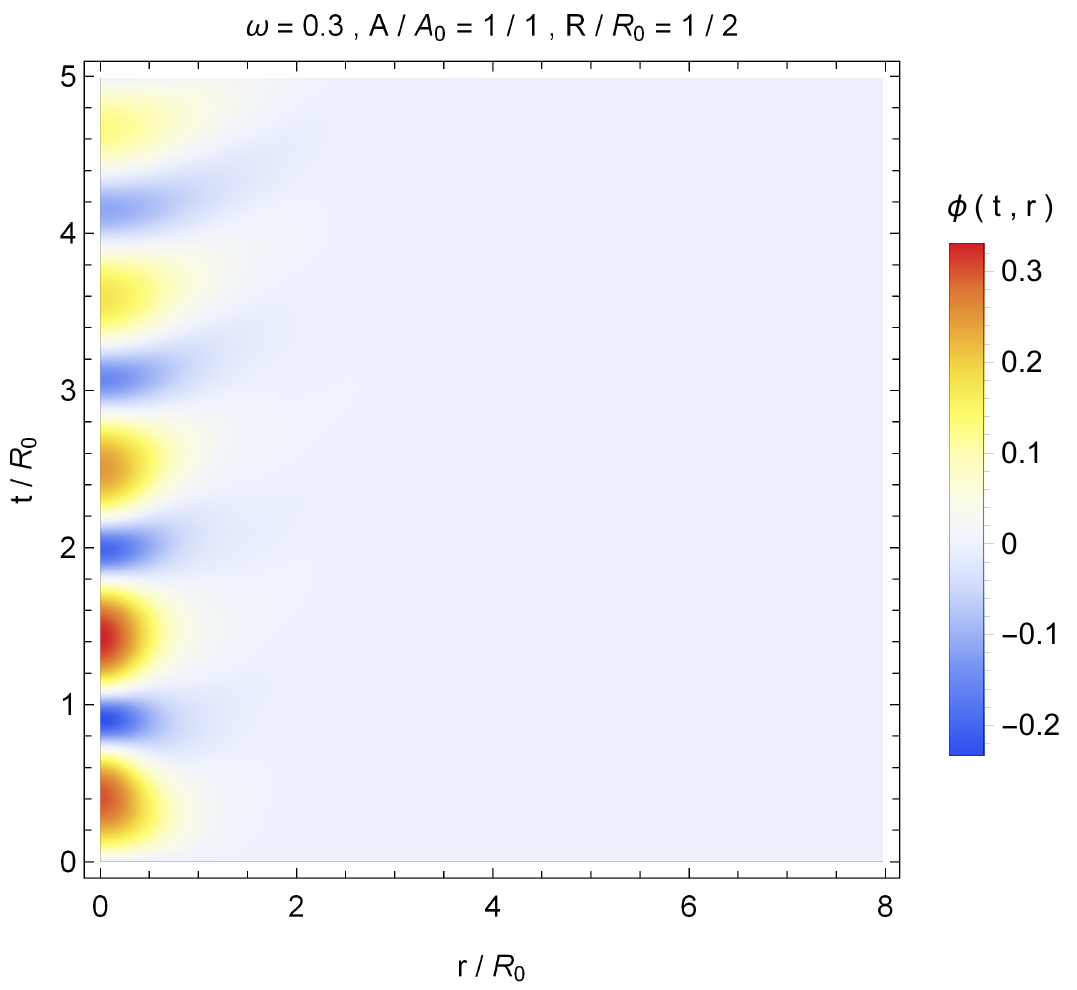}
\includegraphics[width=0.5\textwidth]{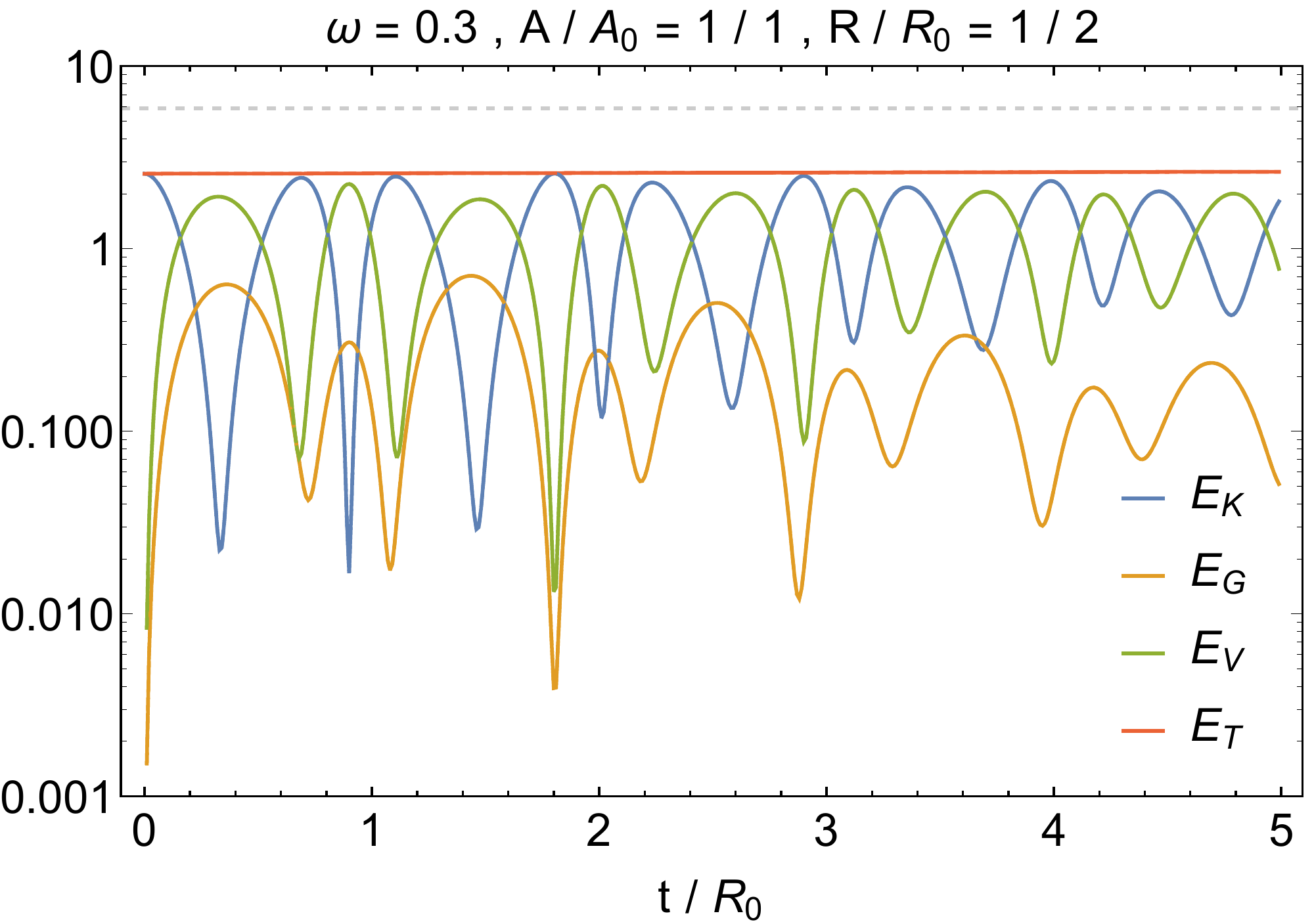}\\
\includegraphics[width=0.4\textwidth]{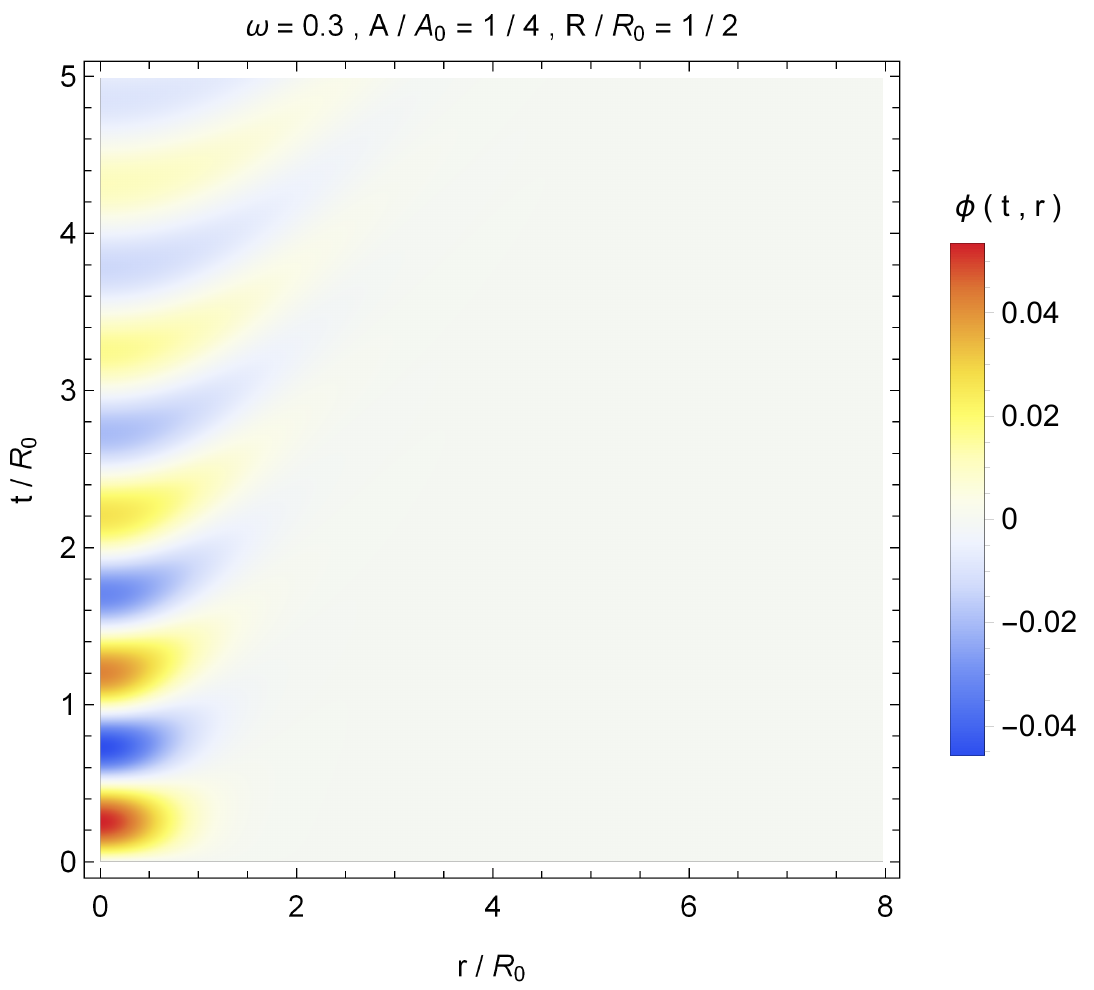}
\includegraphics[width=0.5\textwidth]{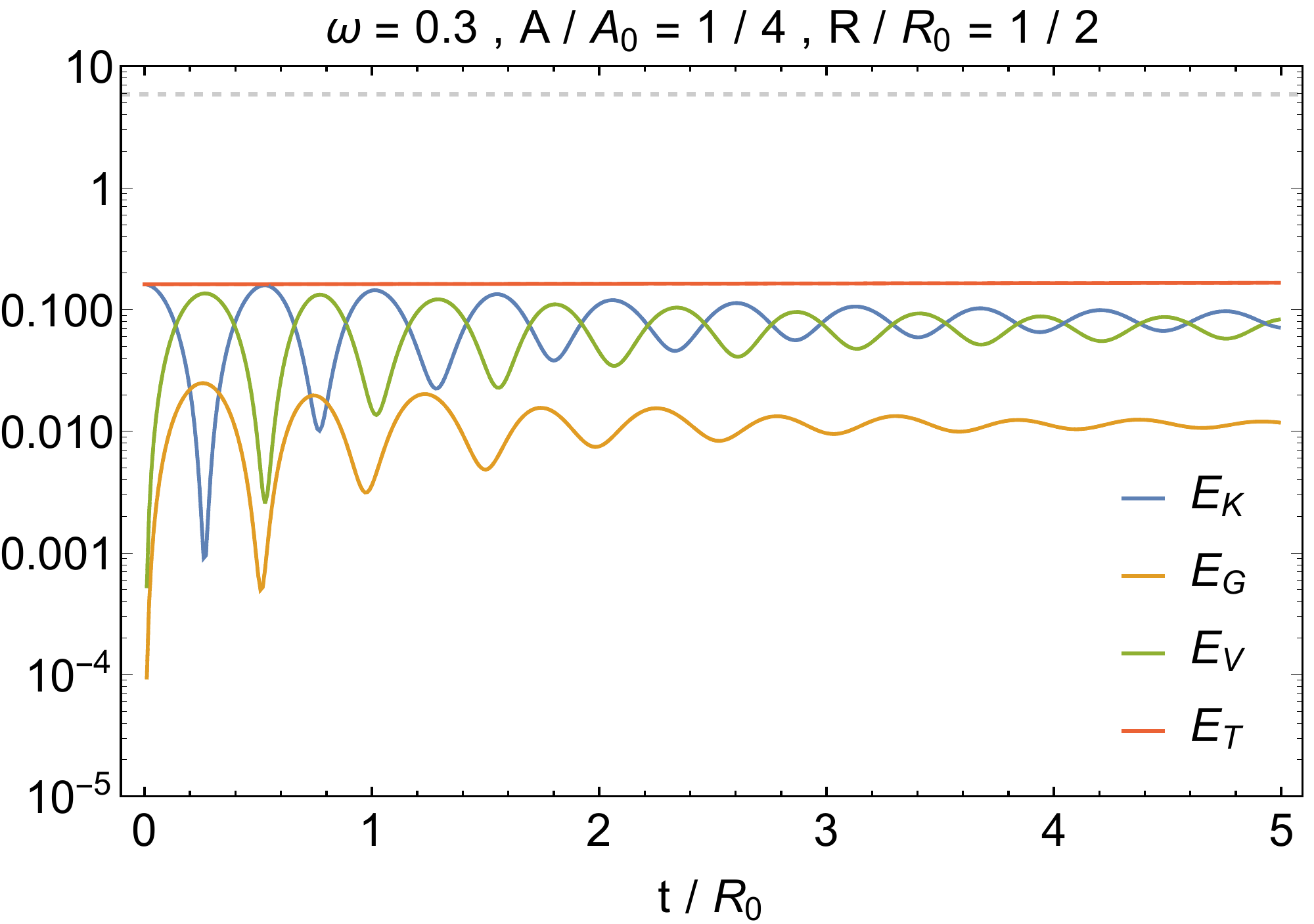}\\
\caption{Three illustrative examples of parameter choices without the appearance of expanding bubble. The time evolutions of field value and the integrated energies are presented in the left and right columns, respectively.}\label{fig:nodecay}
\end{figure*}

If the initial total kinetic energy is smaller than the critical energy threshold specified above, the field value (left column) will simply oscillate around the false vacuum minimum with nearly constant (first example), decreasing (second example), or damping (third example) amplitudes for the time evolutions of total gradient energy (right column), as shown with three illustrative examples in Fig. \ref{fig:nodecay}. This is the case shown with orange shading in the right panel of Fig. \ref{fig:paraspace}. As an aside, the general feature of damping/decreasing amplitude for the time evolution of total gradient energy indicates that the necessary condition for such a semiclassical vacuum decay could also be found by identifying the growing modes of total gradient energy. 

\section{Conclusion and discussions}\label{sec:conclusion}

The flyover vacuum decay is recently proposed as a specific channel of semiclassical vacuum decay, which is realized with an initial Gaussian profile in time derivative of a homogeneous field that is initially sitting at false vacuum. The flyover vacuum decay converts the false vacuum region into the true one but purely along the classical evolutions of the initial condition that is large enough to classically flyover the potential barrier. We generalize the flyover vacuum decay channel by including those initial Gaussian profiles of field velocity that are nowhere larger than the critical velocity to overcome the barrier directly for each local degree of freedom. We have shown by scanning over the whole parameter space of initial condition with numerical simulations that the dubbed pop-up vacuum decay could be possible as long as the initial kinetic energy of the total simulation region is larger than some critical energy threshold, of which an approximated form is estimated and evaluated for the thick-wall case. The thin-wall case should also have the same form but with different numerical prefactor. 

Several improvements could be made as follows: first, a precise determination of such a critical energy threshold should be carried out analytically, possibly by studying the growing modes of the total gradient energy. Second, more general initial configurations with fluctuations in both field value and field velocity should be explored with corresponding necessary condition to be determined for such a semiclassical vacuum decay. Third, similar to the flyover vacuum decay, the gravity effect should also be included for the general semiclassical vacuum decay.

\begin{acknowledgments}
The numerical simulation is implemented in this paper by \texttt{Mathematica 11.3.0}. I would like to thank Heling Deng for the help of numerical simulation carried out originally by \texttt{C}++. I am grateful to Jose J. Blanco-Pillado, Laurence Ford, Mark Hertzberg, Matthew Johnson, Ken Olum, Alexander Vilenkin and Masaki Yamada for the stimulating discussions during many of the lunch time. S. -J. W. is supported by the postdoctoral scholarship of Tufts University from NSF.
\end{acknowledgments}

\bibliographystyle{utphys}
\bibliography{ref}

\end{document}